\documentclass[%
reprint,amsmath,amssymb,
aps,superscriptaddress]{revtex4-2}

\usepackage{graphicx}
\usepackage{dcolumn}
\usepackage{bm}

\usepackage{epsfig}
\usepackage[english]{babel}
\usepackage{latexsym}
\usepackage{graphics}
\usepackage{subfigure}
\usepackage{graphicx}
\usepackage{amsmath}
\usepackage{hyperref}
\usepackage{amssymb}
\usepackage{color}
\usepackage{caption3}

\begin{document}
	\title{A Coulomb-included model for high-order harmonic generation from atoms}
	
	\author{Y. G. Peng}
	\homepage{These authors contribute equally to this paper.}
	\affiliation{College of Physics and Hebei Key Laboratory of Photophysics Research and Application and Physics Postdoctoral Research Station, Hebei Normal University, Shijiazhuang, China}
	\affiliation{College of Physics and Information Technology and Quantum Materials and Devices Key Laboratory of Shaanxi Province's High Education Institution, Shaan'xi Normal University, Xi'an, China}
	\author{J. Y. Che}
	\homepage{These authors contribute equally to this paper.}
	\affiliation{College of Physics, Henan Normal University, Xinxiang, China}
	\author{R. H. Xu}
	\affiliation{Institute of Applied Physics and Computational Mathematics, Beijing, China}
	\author{S. Wang}
	\email{phywangshang@163.com}
	\affiliation{College of Physics and Hebei Key Laboratory of Photophysics Research and Application and Physics Postdoctoral Research Station, Hebei Normal University, Shijiazhuang, China}
	\author{X. J. Xie}
	\email{xiexuejiaohn@163.com}
	\affiliation{College of Physics and Electronic Information, Luoyang Normal University, Luoyang, China}
	\author{Y. J. Chen}
	\email{chenyjhb@gmail.com}
	\affiliation{College of Physics and Information Technology and Quantum Materials and Devices Key Laboratory of Shaanxi Province's High Education Institution, Shaan'xi Normal University, Xi'an, China}
	
	\date{\today}
	
	\begin{abstract}
		In strong laser-atom interactions, the Coulomb potential can affect the trajectories of rescattering electron in high-order harmonic generation (HHG). Here, by constructing a semi-analytical Coulomb-included model and comparing it with numerical experiments that allow for direct observation of electron trajectories, we identify the role of Coulomb potential in different processes of HHG. We show that the symmetry of the system determined by Coulomb potential plays an important role in the ionization process of HHG, inducing the tunneling-out time of electrons to shift towards earlier times. This symmetry-related effect reflects the quantum properties of atomic systems, in sharp contrast to the classical Coulomb-induced acceleration in the recombination process. In particular, compared with other strong-filed models, the scaling law of the amplitude of HHG electron trajectories predicted by this model agrees with the numerical experiments, indicating that the model developed here can be used to quantitatively describe HHG. This model can also be used to study strong-field ionization significantly influenced by rescattering. 		
	\end{abstract}
	\maketitle
	
	\section{Introduction}
	The interaction between strong laser fields and atoms or molecules has attracted widespread attention \cite{F. Krausz}. When the intensity of the laser is high enough so that the electric field force is comparable to the Coulomb force, the Coulomb potential is bent by the electrical field, forming a barrier. The valence electron within the atom or molecule can escape from the barrier through tunneling. The tunneling electron can be ionized by the laser field directly. It can also be driven back to the parent ion and undergo a rescattering event. The rescattering electron can recombine with the parent ion with the emission of high-order harmonics, or recollide with the parent ion with inducing a double-ionization event, etc.. These tunneling triggering processes have been termed as above-threshold ionization(ATI) \cite{Kulander1,Kulander2,M. Lewenstein}, high-order harmonic generation (HHG) \cite{P. B. Corkum,M. Lewenstein2,M. Lewenstein3}, and non-sequence double ionization(NSDI)  \cite{Becker W}, respectively.  
	
	The HHG has a wide range of application in ultrafast science. For example, the HHG can be used to generate attosecond pulses \cite{A.L'Huillier,p.Agostini,Hentschel}, which provide the possibility for ultrafast measurement of attosecond time resolution. The HHG can also be used as a tool to probe the structure \cite{D. Zeidler,S. Haessler,M. Negro,chen} and  dynamics \cite{M. Lein1,O. Smirnova,B. K. McFarland,A. D. Shiner} of electrons within molecules. Theoretically, the HHG process can be described by the classical three-step model (CM) \cite{P. B. Corkum}. According to the CM, the rescattering electron responsible for HHG experiences these three steps of tunneling, propagation and recombination. The HHG can also be described by the strong-field approximation (SFA) \cite{M. Lewenstein2,M. Lewenstein3} which can be understood as the quantum-version three-step model. In SFA, these three steps of HHG are described quantum mechanically. 
	
	Under the framework of CM and SFA, the HHG process can be further understood in terms of trajectories of rescattering electrons \cite{P. B. Corkum,M. Lewenstein3}. The electron trajectories are characterized by the ionization (tunneling-out) time, the recombination (return) time and the return energy of the rescattering electron. Within half a laser cycle, two electron trajectories are generated that contribute to the emission of a harmonic with specific energy: one with an earlier tunneling-out time and a later return time, known as the long trajectory, and the other with a later tunneling-out time and an earlier return time, known as the short trajectory. These electron trajectories provide a mapping between time and experimental observables. Using the mapping, one can reconstruct the ionization time and recombination time of HHG through the experimental observables \cite{M. Y. Ivanov,Lein2013,L. Torlina,Hongchuan Du}.
	
	Recent studies show that the tunneling-out time and the return time of electron trajectories can be influenced by some factors, such as the Coulomb effect \cite{L. Torlina,Hongchuan Du,Hongchuan Du2,xie2,xie1} and the electron wave packet deformation \cite{S. Qiao}. Due to this influence, the tunneling-out time at which the tunneling electron exits the barrier becomes to differ from the ionization time at which the rescattering electron is free,  while for the SFA and CM where the Coulomb potential is neglected, these two times agree with each other. In order to explore the Coulomb effect on the HHG electron trajectory, some strong-field models including Coulomb potential have been developed, such as the analytical R-matrix method (ARM) \cite{L. Torlina} and the Coulomb-modified SFA (MSFA) \cite{xie2,xie1}. These models consider the role of Coulomb potential in HHG electron trajectories from different perspectives. Directly verifying the accuracy of HHG electron trajectories predicted by theory models is a difficult task, as these electron trajectories cannot be directly measured in experiments. Fortunately, the return time of HHG electron trajectories can be directly extracted through results of numerical experiments using the Gabor analysis \cite{M. Lein2}, which provides information on the trajectory of rescattering electrons after recombination with the parent ion, or the wave-packet-tracing procedure \cite{Tong2}, which can track the trajectory of rescattering electrons before the recombination. However, the tunneling-out time of HHG electron trajectories still cannot be directly obtained. Recent studies have shown that the tunneling-out time of electron trajectories is closely related to the amplitude of electron trajectories \cite{xie2}. Therefore, the amplitude of electron trajectories obtained from numerical experiments can be regarded as a numerical observable value to test the accuracy of the tunneling-out time predicted by different models. The results obtained by MSFA in \cite{xie2,xie1} indicate that the tunneling-out time of HHG electron trajectories is several tens of attoseconds earlier than that predicted by SFA. This earlier tunneling time leads to a significant increase in the amplitude of tunneling electrons, resulting in a significant increase in the amplitude of long and short HHG electron trajectories, especially for short trajectories. Although the MSFA can provide some insights into effects of Coulomb potential on HHG, it is a numerical model which considers the Coulomb potential through numerical solution of Newton equation with the initial conditions given by the SFA. Therefore, the MSFA cannot provide a detailed mechanism or clear picture of how the long-range Coulomb potential influences the HHG process. Particularly, the MSFA shows some fundamental limitations in quantitatively describing the Coulomb effect on the ionization process (see \cite{xie3} and the Appendix). Due to the above reasons, further efforts are needed to develop a strong-field model that can accurately and analytically describe effects of Coulomb potential on HHG electron trajectories in a wide parameter region.
	
	Recently, the tunnel-response-classical-motion (TRCM) model is proposed to describe tunneling ionization of atoms \cite{xie3,Huang2023,Jia-Nan Wu} and  molecules \cite{Peng,Shen}. The TRCM arises from Coulomb-included trajectory-based strong-field models \cite{Goreslavski2004,Yan2010,Li2014,Lai2015,Shvetsov2016}. Different from other Coulomb-included models for describing strong-field ionization \cite{xie2,xie1,Goreslavski2004,Yan2010,Li2014,Lai2015,Shvetsov2016,Bray2018,Peng2025} which emphasize the far-nucleus Coulomb effect, the TRCM puts an emphasis on the near-nucleus Coulomb effect. It treats the near-nucleus Coulomb effect analytically and semiclassically around the tunnel exit and is able to quantitatively explain some complex ATI phenomena. For example, in \cite{xie3,cheaps,Huang2023}, it is shown that the TRCM can reproduce the experimental curves of attoclock for different targets and laser parameters \cite{Sainadh,Boge,Landsman,Eckle3,Quan,Torlina2015,Landsman2013}. The TRCM can also analytically derive the scaling law of attoclock observable values relative to the laser parameters of an elliptical laser field  \cite{Huang2023}, give an analytical description of the characteristic structure of photoelectron momentum distribution (PMD) in an orthogonal two-color laser field \cite{Jia-Nan Wu}, and analytically evaluate the effect of internuclear distance on PMDs of small molecules \cite{Peng} as well as the effect of charge resonance and four-body interaction on PMDs of stretched molecular ions \cite{Shen}, providing a simple Coulomb-included physical picture for tunneling ionization of atoms and molecules in strong laser fields. The TRCM does not consider the Coulomb potential after tunneling and therefore cannot describe the Coulomb effect on the rescattering (recombination) process of HHG. Very recently, the basic assumption in the TRCM related to the Coulomb-related symmetry of the tunneling electron wave packet in strong-field ionization is validated in \cite{chen2025,Wu2025}.
	
	In this paper, by classically considering the Coulomb potential when the rescattering electron moves towards the nucleus, we generalize the TRCM to describe the HHG process. We first solve the time-dependent Schr\"{o}dinger equation (TDSE) numerically and obtain the amplitude and return time of electron trajectories from TDSE wave function with the wave-packet-tracing procedure. Then we compare the predictions of models with TDSE results. Compared to other models, the generalized TRCM provides results consistent with TDSE. It also gives a clear Coulomb-included physical picture for the tunneling ionization process of HHG. Specifically,  the tunneling-out time of HHG electron trajectories predicted by this model is several tens of attoseconds earlier than that of MSFA, resulting in larger amplitudes of short electron trajectories and closer results to TDSE. This earlier tunneling-out time is closely related to the Coulomb-induced ionization time lag $\tau$, which can be evaluated by a concise expression composed of basic laser and atomic parameters in the generalized model. This time lag $\tau$ is related to an initial velocity, which is induced by the Coulomb potential near the atomic nucleus, at the tunnel exit and prevents the tunneling electron from escaping. The value of this velocity can be evaluated using the virial theorem which emphasizes the symmetry of the system related to the Coulomb potential and the direction of this velocity is opposite to that of tunneling. The tunneling electron requires a short period of time $\tau$ to obtain an impulse from the laser field to overcome the initial Coulomb-induced velocity, and then the tunneling electron is ionized. The existence of a time lag $\tau$ between tunneling-out time and ionization time allows for such HHG electron trajectories, where the tunneling-out time is located in the rising edge of the laser electric field. In SFA where the tunneling-out time is equal to the ionization time, such trajectories are prohibited. By contrast, the MSFA can also predict such trajectories, but due to the complex motion of the tunneling-out electron in laser and Coulomb fields, it is difficult to define the ionization time and the potential mechanism is also difficult to identify. The return time of HHG electron trajectories predicted by the generalized TRCM is similar to that of the MSFA and earlier than that predicted by the SFA, indicating the important role of Coulomb acceleration in the recombination process of HHG. In particular, when the scaling law for the dependence of the amplitude of short and long electron trajectories on the return time predicted by MSFA differs remarkably from TDSE, the prediction of TRCM is consistent with TDSE under different laser parameters. Since the scaling law reflects the most fundamental dynamical characteristics of the HHG process, this agreement suggests the applicability of the TRCM model developed here for quantitatively describing the HHG electron trajectories.	
	
	\section{Theoretical Methods}
	\subsection{ TDSE}
	We first choose the H atom as the target which can be simulated accurately. We solve the three-dimensional (3D) TDSE to describe the interaction of the H atom with a strong laser filed using the generalized pseudospectral method \cite{Tong}. In the length gauge, the Hamiltonian of the system studied can be written as $H(t)=H_{0}+\mathbf{E}(t)\cdot\mathbf{r}$ (in atomic unit of $\hbar =e={m}_{e}=1$). Here, $H_{0}=\mathbf{p}^{2}/2+V(\mathbf{r})$ is the field-free Hamiltonian, $V(\mathbf{r})=-Z/r$ is the Coulomb potential, $Z=1$ is the nuclear charge and $\textbf{r}$ is the electronic coordinate. The term $\mathbf{E}(t)=\mathbf{e}_xf(t)E_{0}\sin(\omega_{0}t)$ is the laser electric field, where $\textbf{e}_x$ is the unit vector along the $x$ axis, $f(t)$ is the envelop function, $E_{0}$ is the amplitude of the laser field, and $\omega_{0}$ is the laser frequency. In our simulations, we use trapezoidal laser pulses with a total duration of ten optical cycles and linear ramp of three optical cycles. After each time step, the wave function $\psi(\mathbf{r},t)$ of TDSE is multiplied by a mask function $F(r)=\cos^{1/4}[\pi(r-r_{0})/2(r_{m}-r_{0})]$ to filter the continuum wave function at the boundary. Here, $r_{0}$ is the absorbing boundary and $r_{m}=400$ a.u. is the grid size \cite{R H XU}.

	To obtain information of return time and amplitude of electron trajectories from numerical experiments, we use the procedure introduced in \cite{Tong2}. We first project the TDSE wave function, which is obtained by an integration equation at a half-cycle period, onto the continuum eigenstates of $H_{0}$ in the inner region when the wave packet returns to the parent core at each time step. Then we obtain the time-energy distribution of the rescattering electron wave packet before recombination. By searching for the local maximum amplitude and the corresponding time of this distribution \cite{Y. J. Chen1}, we obtain the amplitude and return time of electron trajectories including short and long HHG trajectories. In the following, we will compare the numerical results with model predictions. The above procedure for obtaining the trajectories of rescattering electron can also be used to study the HHG electron trajectories of multiple return \cite{R H XU} and trajectories of small molecules \cite{Y. J. Chen2}.
	
	\subsection{ SFA}
	In the general SFA, the electron trajectories of HHG are given by the following saddle-point equations \cite{M. Lewenstein2,M. Lewenstein3}:
	\begin{align}
		[\textbf{p}_{st}+\textbf{A}(t_s)]^2/2=-I_p,\label{eq:1}\\
		[\textbf{p}_{st}+\textbf{A}(t_{r}')]^2/2=\Omega -I_p. \label{eq:2}
	\end{align}
	Here, $\textbf{p}_{st}\equiv{\int_{t_{s}}^{t_{r}'}\textbf{A}(t')dt'}/{(t_{r}'-t_{s})} $ is the saddle-point momentum, $\textbf{A}(t)=-\int^{t}{\textbf{E}(t')}dt'$ is the vector potential of electric field $\textbf{E}(t)$, $\Omega$ is the energy of high-order harmonics and $I_p$ is the ionization energy of the system.  The solutions $(t_s,t_r')$ of the above equations are complex. The real part $t_0$ of the complex solution $t_s$ is called the tunneling-out time, at which the electron tunnels out of the barrier formed by the laser field and the atomic potential. In the SFA which does not consider the Coulomb potential, the tunneling-out time $t_0$ is also the ionization time at which the tunneling electron becomes a free particle. However, when the Coulomb potential is considered, the situation is different \cite{xie3}. We will discuss this point in detail later. The free electron can be driven back to the nucleus by the laser field and be rescattered by the Coulomb potential. Then, it may recombine with the parent ion and contribute to HHG. For convenience, the free electron which is experiencing a rescattering event is also called the rescattering electron. The real part $t_r$ of the complex solution $t_r'$ is called the return time (or the recombination time) at which the rescattering electron returns to and recombines with the parent ion with the emission of a harmonic $\Omega$. Each pair of solutions $(t_s,t_r')$ is also called the electron trajectory or the quantum orbit (QO) for HHG \cite{M. Lewenstein3}, including the short trajectories with the excursion time $t_f=t_r-t_0$ shorter than half a laser cycle, the long trajectories with $t_f$ near to a laser cycle, and multiple returns with $t_f$ longer than a laser cycle. Once the electron trajectories are obtained, the amplitude of the rescattering electron, which is born at the time $t_0$ and return to the nucleus at the time $t_r$ can be approximately evaluated with the following expression \cite{xie2}:
	\begin{equation}
		F_{S}(t_{s},t_r',\Omega)\equiv F_{S}(t_{0},t_{r},\Omega)\propto(1/{t_f})^{1.5}e^{b'}.
	\end{equation}
	
	In the above expression, the factor $(1/{t_{f}})^{1.5}$ represents the quantum diffusion of the rescattering electron wave pack when it propagates in the laser field, and the factor $e^{b'}$ can be understood as the amplitude of the rescattering electron wave packet at time $t_0$, which is first formed by a small portion of the initial electron-state wave packet tunneling out of the potential barrier at that time $t_0$. Here, $b'$ is the imaginary part of the action $S'(t_s,t_r',\Omega)$ with the expression of
	\begin{equation}
		S'(t_s,t_r',\Omega)=\int_{t_{s}}^{t_r'}\{{[\textbf{p}_{st}+\textbf{A}(t)]^2/2}+I_{p}\}dt-\Omega t_r'.
	\end{equation}
	The use of Eq. (3) allows us to directly compare the HHG electron trajectories predicted by different theoretical models, not only for the return time, but also for the amplitude of electron trajectories. We will also discuss this point below Eq. (11).
	
	\subsection{ MSFA}
	Although the SFA gives an appropriate description of HHG process \cite{M. Lewenstein2,M. Lewenstein3}, it neglects the long-range Coulomb potential. To study the coulomb effect on HHG electron trajectories, a Coulomb-modified model which is based on SFA and saddle-point theory and termed as MSFA has been developed \cite{xie2}. To obtain the Coulomb-modified HHG electron trajectories in MSFA, one can first solve the following saddle-point equation for ATI  \cite{M. Lewenstein}:
	\begin{equation}
		[\textbf{p}+\textbf{A}(t_s)]^2/{2}=-I_p.
	\end{equation}
	Here, $\textbf{p}$ is the drift momentum. The solution  $t_s=t_0+it_x$  is complex and the real part $t_0$ of $t_s$ corresponds to the tunnleing-out time. The amplitude $b(\textbf{p},t_{s})$ of the photoelectron with the momentum $\textbf{p}$ is given by the expression of  $b(\textbf{p},t_{s})\equiv b(\textbf{p},t_{0})\propto e^{b}$. Here, b is the imaginary part of the following action $S(\textbf{p},t_{s})$ for ATI \cite{M. Lewenstein}
	\begin{equation}
		S(\textbf{p},t_{s})=\int_{t_{s}}\{[\textbf{p}+\textbf{A}(t)]^2/2+I_{p}\}dt.
	\end{equation}
	Then, the Coulomb-modified electron trajectories for HHG are obtained by solving the following Newton equation
	\begin{equation}
		\ddot{\textbf{r}}(\textbf{p},t)=-\textbf{E}(t)-{{\nabla }_{\mathbf{r}}}V(\mathbf{r}(\textbf{p},t)),
	\end{equation}
	for each electron trajectory $(\textbf{p},t_{s})$ , with the initial conditions given by Eq. (5). Here, $V(\mathbf{r})$ is the Coulomb potential.  Specifically, the initial velocity (i.e, the exit velocity) is
	\begin{equation}
		\dot{\textbf{r}}(\textbf{p},t_0)=\textbf{v}(t_0)=\textbf{p}+\textbf{A}(t_0),
	\end{equation}
	and the initial position (i.e, the exit position) is
	\begin{equation}
		{\textbf{r}}(\textbf{p},t_0)=Re(\int_{t_{s}}^{t_0}[\textbf{p}+\textbf{A}(t)]dt).
	\end{equation}
	Then, the instantaneous energy of the ionized electron moving in the laser field and the Coulomb field can be evaluated with
	\begin{equation}
		E_a(t)=\dot{\textbf{r}}^2(\textbf{p},t)/2+V(\mathbf{r}(\textbf{p},t)).
	\end{equation}
	
	We solve Eq. (7) using the Runge-Kutta method with variable-step integration. In our simulations with Eq. (7), the return time $t_r$ of the rescattering electron with the momentum $\textbf{p}$ and the birth time $t_0$, to the parent ion is defined as the time at which the relation of ${\left| \textbf{r} (\textbf{p},t_r)\right|}\le 0.1$ a.u. with $t_r\textgreater t_0$ is satisfied. Then, the Coulomb-modified HHG trajectories $(t_0,t_r)$ for the emission of a harmonic $\Omega=E_a(t_r)+I_p$ are obtained. In MSFA, the amplitude of the rescattering electron related to the trajectory $(t_0,t_r)$  and the harmonic $\Omega$ can be evaluated with the following expression \cite{xie2,xie1}:
	\begin{equation}
		F_{M}(t_{s},t_{r},\Omega)\equiv F_{M}(t_{0},t_{r},\Omega) \propto(1/{t_f})^{1.5}e^{b}.
	\end{equation}
	Here, $t_f=t_r-t_{0}$ is the excursion time of the rescattering electron and $b$ is the imaginary part of the action $S(\textbf{p},t_{s})$ of Eq. (6). It should be noted that the term $e^{b}$ in Eq. (11) of MSFA is only related to the tunneling-out time obtained from the saddle-point equation of Eq. (5), whereas the term $ e^{b'}$ in Eq. (3) of SFA is related to both the tunneling-out time and the return time obtained from the saddle-point equations of Eqs. (1) and (2). However, as shown in Ref. \cite{xie1}, for the long trajectories, the amplitude of Eq. (11) is near to the amplitude of Eq. (3) and the amplitude obtained from numerical experiments. The amplitudes of MSFA and SFA differ remarkably from each other for the short trajectories and the prediction of MSFA that considers the Coulomb effect is closer to the TDSE result. In particular, for a short-range potential, the results of Eq. (11) almost coincide with the results of Eq. (3) for both long and short trajectories. 
	
	Although MSFA predictions agree better with TDSE results than SFA predictions \cite{xie2,xie1}, the MSFA is a numerical model. It can not provide a clear picture or a detailed mechanism of how the long-range Coulomb potential affects the HHG electron trajectories. The MSFA attributes the complex Coulomb effect in tunneling ionization to a simple Coulomb-induced ionization time lag and indicates that this lag plays an important role in the subsequent dynamic of electrons after tunneling. However, it can not give a clear definition and therefore can not provide an analytical expression for the time lag.

	\subsection{ TRCM}
	Based on the SFA, recently, a Coulomb-included strong-field model termed as TRCM is further developed for the ATI \cite{xie3}. Similar to the MSFA, the TRCM considers the Coulomb correction for each SFA electron trajectory $(\textbf{p},t_0)$. As this correction in MSFA is performed through numerical solution of Newton equation including both electric-field and Coulomb forces after the tunneling electron exits the barrier, this correction in TRCM is treated analytically by considering the Coulomb-related symmetry of the tunneling electron wave packet at the tunnel exit. With the analytical treatment, the TRCM is able to give a clear definition and a concise expression for the ionization time lag $\tau$ induced by the Coulomb potential. Specifically, in the TRCM, it is assumed that due to the existence of the Coulomb potential with the central symmetry for an actual atom, at the tunnel exit $\textbf{r}_0$, the tunneling electron is still located at a quasi-bound state which approximately satisfies the virial theorem \cite{xie3}. This assumption related to the virial theorem indicates that during the process of strong-field ionization, the atomic system tends to keep its basic symmetry as much as possible before an ionization event occurs. According to this assumption which has been further verified in a recent work \cite{chen2025}, the average potential energy of the quasi-bound state is $\left. \left\langle V(\textbf{r}) \right. \right\rangle \approx V({\textbf{r}_{0}})$ and the average kinetic energy is $n_f\textbf{v}^2 _{i}/2=\left. \left\langle \textbf{v}^2 /2 \right. \right\rangle \approx -\left. \left\langle V(\textbf{r}) \right. \right\rangle/2 \approx -V({{\textbf{r}}_{0}})/2$. Here, $n_f$ is the dimension of the system studied. For example, for the 3D cases mainly explored in the paper, we have $n_f=3$. The direction of the velocity $\textbf{v}_i$ is opposite to the position vector $\textbf{r}_0$, namely, $\textbf{v}_i= -{v}_i\textbf{r}_0/{r}_0$, and the amplitude  ${v}_i =\left| \textbf{v}_i \right|$  is given by the following expression:
	\begin{equation}
		{v}_{i}=\sqrt{\left| V(\textbf{r}_0) \right|/{n_{f}}}.
	\end{equation}
	Here, $\textbf{r}_0\equiv \textbf{r}(t_0)\equiv \textbf{r}(\textbf{p}, t_0)$ is given by  Eq. (9). The coulomb-induced velocity $\textbf{v}_{i}$ reflects the basic symmetry requirement of the Coulomb potential on the electronic state. 
	
	In the SFA, the drift momentum $\textbf{p}$ is relevant to the tunneling-out (ionization) time $t_0$, and the mapping between $\textbf{p}$ and $t_0$ is
	\begin{equation}
		\textbf{p}=\textbf{v}(t_0)-\textbf{A}(t_0).
	\end{equation}
	In the TRCM, when the coulomb potential is considered, the drift momentum changes from $\textbf{p}$ to $\textbf{p}'$. Considering the Coulomb-induced velocity $\textbf{v}_i$ at the tunneling-out time $t_0$, the mapping between $\textbf{p}'$ and  $t_0$ is
	\begin{equation}
		\textbf{p}' =\textbf{p}+\textbf{v}_{i}=\textbf{v}(t_0)-\textbf{A}(t_0)+\textbf{v}_{i}.
	\end{equation}
	In the above expression, the existence of the Coulomb-induced velocity $\textbf{v}_i$ implies that at the tunneling-out time $t_0$, the electron is still subject to the Coulomb effect and therefore cannot be considered as a free particle. Accordingly, the time $t_0$ cannot be considered as the ionization time. To overcome the velocity  $\textbf{v}_i$,  a time lag $\tau$ relative to $t_0$, is needed. The value of $\tau$  can be calculated with the following expression \cite{xie3}:
	\begin{equation}
		\tau\approx{v}_{i}/\left| \textbf{E}(t_0)\right|=\sqrt{\left| V(\textbf{r}_0) \right|/n_{f}}/\left| \textbf{E}(t_0)\right|.
	\end{equation}
	With the lag $\tau$, Eq. (14) can be rewritten as
	\begin{equation}
		\textbf{p}'\approx\textbf{v}(t_0)-\textbf{A}(t_i=t_0+\tau).
	\end{equation}
	The above expression gives the mapping between the Coulomb-modified momentum $\textbf{p}'$ and the time  $t_i=t_0+\tau$ which can be considered as the ionization time. It shows that the complex Coulomb effect in tunneling ionization can be attributed to a simple time lag $\tau$ relative to the tunneling-out time $t_0$. After the time $t_i$, the Coulomb effect can be neglected as in the general SFA and the movement of the electron in the laser field can be treated classically. A detailed discussion on the applicability of the TRCM model for ionization developed in \cite{xie3,Huang2023,Jia-Nan Wu,Peng,Shen} can be found in the Appendix.
	
	However, for the rescattering process, in which the rescattering electron will return to the parent ion, the Coulomb effect needs to be reconsidered. This rescattering is important for many strong-field processes such as HHG \cite{P. B. Corkum,M. Lewenstein2,M. Lewenstein3}, high-order ATI \cite{Kulander1,Kulander2,M. Lewenstein} and NSDI \cite{Becker W}. To consider the Coulomb effect on this rescattering, we include the Coulomb potential into the TRCM when the ionized electron driven by the laser field changes its direction and begins to move towards the parent ion. Specifically, instead of Eq. (7), we solve the following equation
	\begin{equation}
		\ddot{\textbf{r}}(\textbf{p}',t)=\left\{\begin{array}{lr}-\textbf{E}(t),&   \,\, t\le t_{e}\\-\textbf{E}(t)-{{\nabla }_{\mathbf{r}}}V(\mathbf{r}(\textbf{p}',t)),& \,\,  t\textgreater t_{e} \end{array}\right.
	\end{equation}
	for each TRCM electron trajectory $(\textbf{p}',t_i)$ originating from the SFA trajectory $(\textbf{p},t_0)$. Here, $t_{e}$ is the time when the instantaneous velocity $\dot{\textbf{r}}(\textbf{p}',t)$ of electron is equal to zero. The initial velocity now used for the evolution of the above Newtonian equation is
	\begin{equation}
		\dot{\textbf{r}}(\textbf{p}',t_i)=\textbf{v}(t_0)=\textbf{p}'+\textbf{A}(t_i)\approx\textbf{p}+\textbf{A}(t_0),
	\end{equation}
	and the initial position used is
	\begin{equation}
		{\textbf{r}}(\textbf{p}',t_i)\approx {\textbf{r}}(\textbf{p},t_0)=Re(\int_{t_{s}}^{t_0}[\textbf{p}+\textbf{A}(t)]dt).
	\end{equation}
	These conditions are similar to those in MSFA, but the initial time is shifted from  $t_0$ to $t_i$, and the corresponding momentum is shifted from $\textbf{p}$ to $\textbf{p}'$. The condition for obtaining the Coulomb-included return time $t_r$ is ${\left| \textbf{r} (\textbf{p}',t_r)\right|}\le 0.1$ a.u. with  $t_r\textgreater t_i$. The energy of harmonic emitted at $t_r$ is $\Omega=E_a(t_r)+I_p$ with $E_a(t_r)=\dot{\textbf{r}}^2(\textbf{p}',t_r)/2+V(\mathbf{r}(\textbf{p}',t_r))$. The amplitude of the rescattering electron related to the trajectory $(t_i,t_r)$ and the harmonic $\Omega$ in the generalized model can be expressed as
	\begin{equation}
		F_{T}(t_{s},t_{r},\Omega)\equiv F_{T}(t_{i},t_{r},\Omega)\propto(1/{t_f})^{1.5}e^{b}.
	\end{equation}
	Here, $t_f=t_r-t_0$ is the excursion time of the rescattering electron in the generalized TRCM relating to Eqs. (17) to (20). It should be emphasized that the generalized TRCM is a semi-analytical model, which provides analytical and quantum descriptions of the Coulomb effect on the tunneling ionization process, and numerical and classical descriptions of the Coulomb effect on the rescattering process. Therefore, it can be used to explore the role of Coulomb potential in different processes of HHG. The amplitude of the Coulomb-included HHG electron trajectory predicted by the generalized TRCM differs remarkably from the prediction of MSFA and is consistent to the TDSE result, as to be shown below. The inclusion of the Coulomb-induced ionization time lag $\tau$ defined by Eq. (15) into the electron trajectory can also give a clear physics picture for the shift of the tunneling-out time of the trajectory towards an earlier time revealed in MSFA simulations. In the following, for simplicity, we will also refer to the generalized TRCM as the TRCM. It should be mentioned that in TRCM, the initial position ${\textbf{r}}(\textbf{p}',t_i)$ at the ionization time $t_i$ is similar to the position ${\textbf{r}}(\textbf{p},t_0)$ at the tunneling-out time $t_0$ \cite{xie3}, suggesting that the quantum diffusion effect closely related to the excursion distance of the rescattering electron can be neglected in the short time interval of $\tau=t_i-t_0$. Therefore, the excursion time $t_f$ in Eq. (20) can also be approximated as $t_f=t_r-t_i$. In MSFA, due to the rapid oscillation of electron trajectories in the combined field of laser electric field and Coulomb field, the ionization time $t_i$ cannot be clearly defined. In addition, in comparison with the general TRCM where the Coulomb potential is ignored after the ionization time $t_i$ \cite{xie3,Huang2023,Jia-Nan Wu,Peng,Shen}, the generalized TRCM developed here considers the Coulomb effect in the rescattering process. Therefore, besides of HHG, it can also be used to explore strong-field ionization where the rescattering plays an important role. More discussions on the significance of the TRCM model for HHG developed here can be found in the Appendix.
	
	Unless mentioned elsewhere, our MSFA and TRCM calculations in the paper are performed using the form of Coulomb potential of $V(\mathbf{r})=-Z/\sqrt{{\mathbf{r}}^{2}+\rho}$ with $Z=\sqrt{2I_p}$ and $\rho=0.5$. The soft-core parameter $\rho$ is used to avoid the singularity in numerical solution of Newton equation. In addition, the parameter ${n_{f}}=3$ is used in Eq. (12) to match the 3D TDSE simulation.
	
	\section{Results and discussions}
	
	\begin{figure}[t]
		\begin{center}
			\rotatebox{0}{\resizebox *{8.5cm}{7.5cm} {\includegraphics {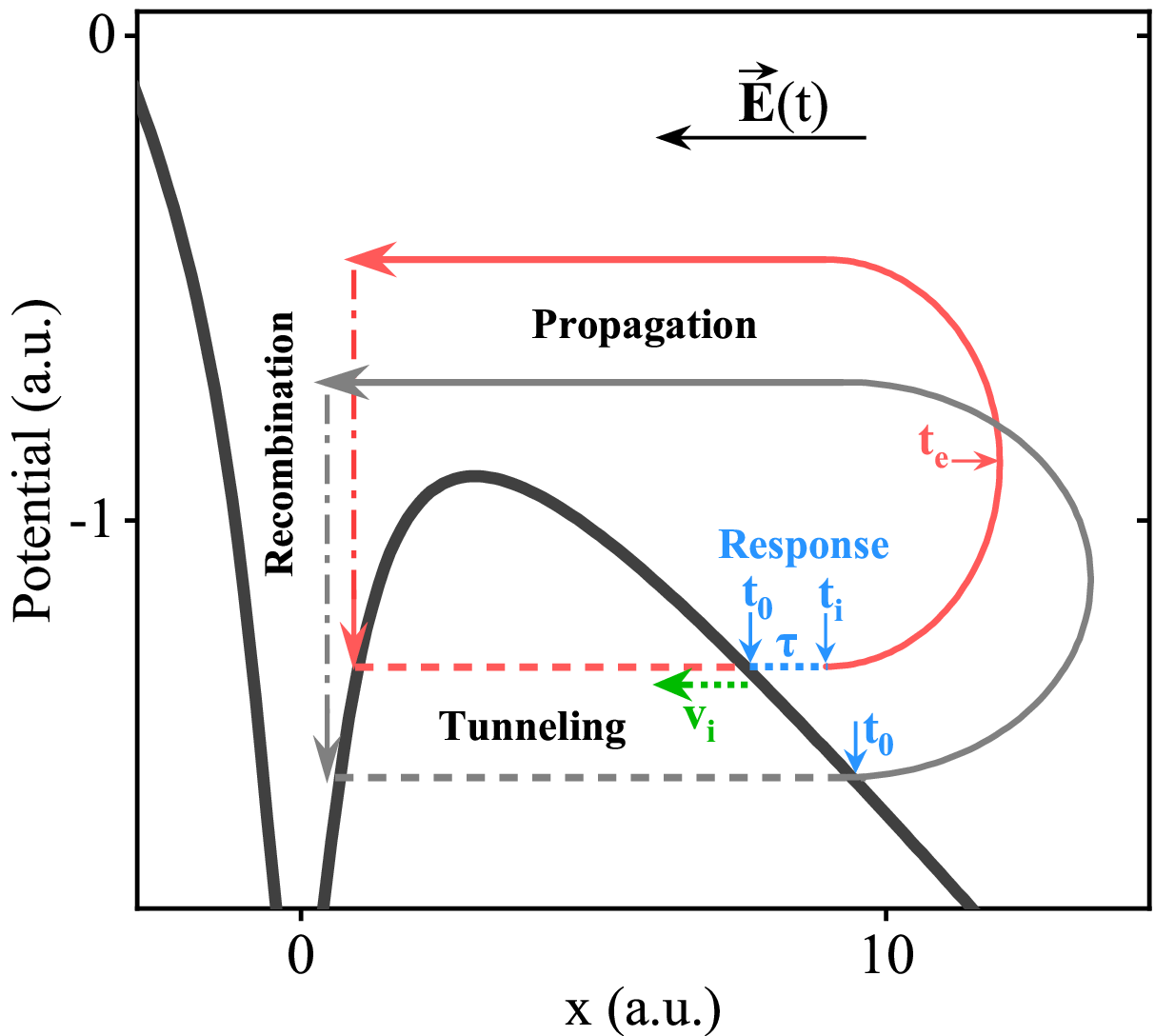}}}
		\end{center}
		\caption{A sketch of the HHG process described by the SFA and the TRCM proposed here. According to the SFA, the HHG is described by a three-step process (the gray curve). That is, the electron firstly tunnels out of the barrier at the time $t_0$; then it propagates in the laser field; when the laser field changes the direction, the electron is driven to return to the nucleus, with the emission of a harmonic $\Omega$. According to the TRCM, the HHG is described by a four-step process of tunneling, response, propagation and recombination (the red curve). Specifically, when considering the Coulomb effect on tunneling, at the tunneling-out time $t_0$, the electron appears at the tunnel exit with a Coulomb-induced nonzero velocity $\textbf{v}_i$. The value of this velocity is determined by the virial theorem that puts an especial emphasis on the symmetry of the system and the direction of this velocity is opposite to the direction of tunneling (the green dotted arrow). A small period of time $\tau$ is needed for the tunneling electron to overcome this velocity. Then the tunneling electron is ionized at the time $t_i=t_0+\tau$ and begins to propagate in the laser field. The additional process between tunneling and propagation is called the response process, as the time lag $\tau$ characterizing this process reflects the basic response time of electrons within the atom to laser-induced photoionization events. When the direction of the electron velocity changes at time $t_e$ implying that a rescattering event occurs, we classically reconsider the coulomb effect in the propagation process, which accelerates the rescattering electron to return to the nucleus.}
		\label{fig:g1}
	\end{figure}

	\begin{figure}[t]
		\begin{center}
			\rotatebox{0}{\resizebox *{8.5cm}{6.5cm} {\includegraphics {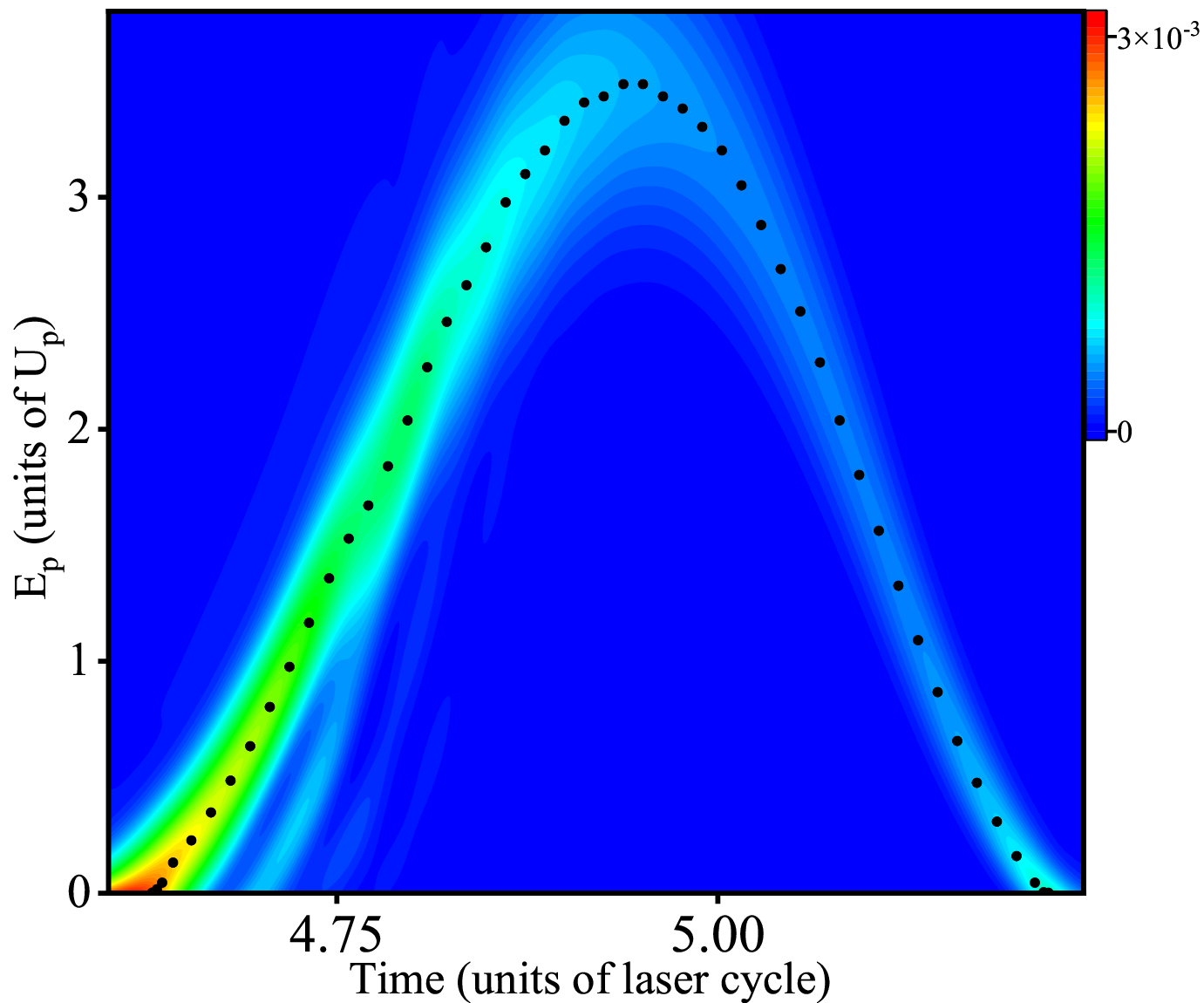}}}
		\end{center}
		\caption{Time-energy distributions of rescattering electron wave packet related to HHG, obtained by 3D TDSE for the H atom. The local maximum amplitudes are marked with black-solid circles. The energy $E_p$ is plotted in the unit of  $U_p=E_0^2/(4\omega_0^2)$. The laser parameters used are $I=2\times10^{14}$W/cm$^{2}$ and $\lambda=900$ nm.}
		\label{fig:g2}
	\end{figure}

	In Fig. \ref{fig:g1}, we display a sketch of the HHG process described by SFA and TRCM. In the SFA without considering the long-rang Coulomb potential, the HHG process can be described by a three-step process of tunneling, propagation and recombination (the gray curve). However, in the TRCM which includes the long-range Coulomb potential, the HHG process can be described by a four-step process of tunneling, response, propagation and recombination (the red curve). The appearance of the response process in TRCM is due to the effect of the Coulomb potential near the atomic nucleus. The response process is related to a Coulomb-induced nonzero velocity at the tunnel exit which prevents ionization of the tunneling electron, as indicated by Eq. (12). Due to the existence of the response process, there is a time lag $\tau=t_i-t_0$ between the tunneling-out time $t_0$ and the ionization time $t_i$ in TRCM, as indicated by Eq. (16). This lag can be analytically evaluated through Eq. (15). It is absent in SFA. As a result, for the emission of a same harmonic $\Omega$, the tunneling-out time $t_0$ predicted by the TRCM will be earlier than that predicted by the SFA. This will further induce the TRCM prediction of HHG amplitude, which depends strongly on the tunneling-out time, to differ from SFA. As the tunneling-out time cannot be directly measured in experiments, in the following discussions, we will compare the predictions of the HHG amplitude closely related to the tunneling-out time between different theory models and TDSE.  Besides the Coulomb effect on the tunneling-out time, the TRCM also considers the Coulomb effect on the return time. This is achieved through numerical solution of Newton equation, including both the electric-field force and the Coulomb force, after the time $t_e$ when the rescattering electron changes its direction of motion and begins to return to the atomic nucleus. In the following discussions, we will also compare the predictions of the return time between different theory models and TDSE. We mention that the MSFA can also predict the response process but can not give the analytical expression of the response time and therefore can not provide the detailed physical mechanism of the response process.
	
	In Fig. \ref{fig:g2}, we show the time-energy distribution (the color coding) of the rescattering electron, obtained by 3D TDSE. From the distribution, we can further obtain the local maximum amplitude (the black solid circle). The curve of the local maximum amplitude gives the HHG electron trajectory of TDSE including the information of return time, return energy and trajectory amplitude. The trajectory amplitude is considered to be exponentially dependent of the tunneling-out time of the electron trajectory and therefore can be used to qualitatively analyze the tunneling-out time  \cite{xie1}. The electron trajectories extracted from TDSE can be directly compared with the predictions of theoretical models, providing numerically observable values to test the accuracy of the model. It should also be noted that the distribution in Fig. \ref{fig:g2} shows two branches corresponding to long and short electron trajectories and the amplitude of the short-trajectory branch with plus chirp is remarkably larger than the long-trajectory one with minus chirp. We will further discuss these characteristics later. 
	
	\begin{figure}[t]
		\begin{center}
			\rotatebox{0}{\resizebox *{8.5cm}{6.5cm} {\includegraphics {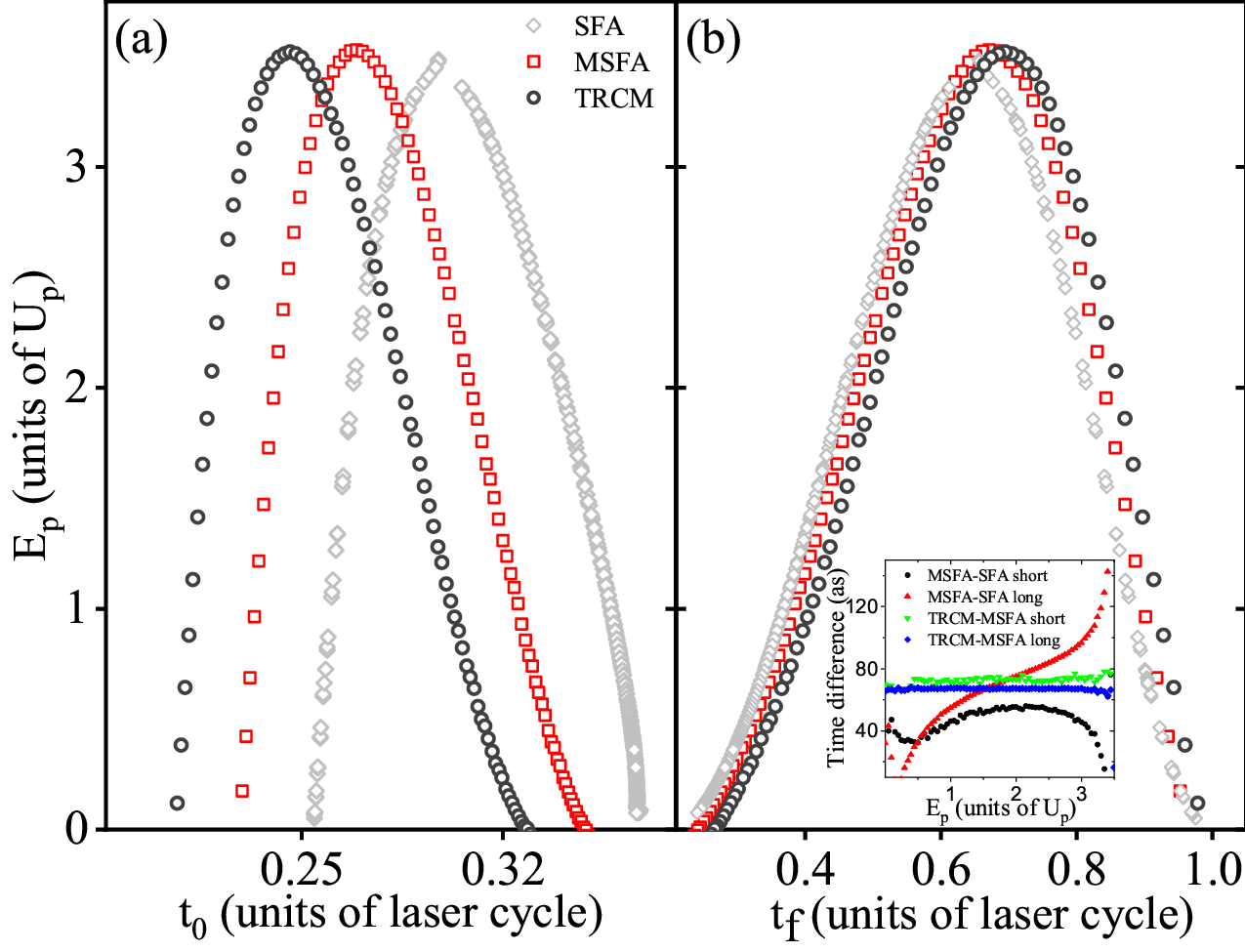}}}
		\end{center}
		\caption{Comparisons of HHG electron trajectories for the H atom, obtained with TRCM, MSFA and SFA. (a) : Tunneling-out time $t_{0}$ versus return energy $E_p$. (b): Excursion time $t_f$ versus return energy $E_p$. The laser parameters used are as in Fig. \ref{fig:g2}. The inset in (b) shows the difference between predictions of different models for the excursion time.}
		\label{fig:g3}
	\end{figure}

	In Fig. \ref{fig:g3}, we compare electron trajectories for H atom obtained with SFA, MSFA, and TRCM. Results of the tunneling-out time $t_0$ versus the return energy $E_p$ are plotted in Fig. \ref{fig:g3}(a), and those of the excursion time $t_f=t_r-t_0$ versus the return energy $E_p$ are plotted in Fig. \ref{fig:g3}(b). Each curve relating to the predictions of a specific model in Fig. \ref{fig:g3} shows two branches. For the case of tunneling-out time in Fig. \ref{fig:g3}(a), the branch with plus chirp corresponds to the long trajectory and that with minus chirp corresponds to the short trajectory. The situation reverses for the case of excursion time in Fig. \ref{fig:g3}(b). One can observe from Fig. \ref{fig:g3}(a), that when the predictions of MSFA for the tunneling-out time are remarkably earlier than those of SFA, the results of TRCM are earlier than the MSFA ones both for the long and short trajectories. Specifically, compared to the MSFA (SFA) curve, the TRCM (MSFA) curve is generally shifted towards an earlier time by about 60 attoseconds. In particular, almost the entire long-trajectory branch of TRCM for different energy is located in the time region of $0.2T<t<0.25T$ corresponding to the rising edge of the laser field. The short-trajectory branch of TRCM is also very near to the peak time $t=0.25T$ of the laser field. Here, $T=2\pi/\omega_0$ is the laser cycle. For SFA, the tunneling-out time for both of long and short trajectories is located in the falling edge of the laser field and the tunneling-out time of long trajectory is nearer to the peak time than the short one, as seen in Fig. \ref{fig:g3}(a). The earlier tunneling-out time predicted by the TRCM arises from the Coulomb-induced ionization time lag $\tau$ (the response time). Due to this lag, electrons tunneling out of the potential barrier at time $t_0$ can remain near the tunnel exit before ionization time $t_i=t_0+\tau$ \cite{xie3,Huang2023}. This allows electron trajectories with tunneling-out times located in the rising edge of the laser field to still contribute to HHG, however, these trajectories are prohibited in SFA. 
	
	For the case of the excursion time in Fig. \ref{fig:g3}(b),  a careful analysis shows there are also somewhat differences in the predictions of these theoretical models. Specifically, by calculating the difference between curves obtained by different methods in Fig. \ref{fig:g3}(b), it can be seen from the inset that the predictions of SFA (MSFA) are, on average, about 70 attoseconds shorter than those of MSFA (TRCM). In other words, the TRCM predicts a longer excursion time than the MSFA and the SFA. However, considering the large time scale of the excursion time (ranging from $0.23T$ to $1T$ with $T\approx124$ a.u. in the current situation), these differences can be ignored, and we can consider the predictions of the models for the excursion time to be comparable. The tunneling-out time $t_{0}$ and the excursion time $t_{f}$ discussed above play a vital role in the amplitude $F$ of HHG electron trajectories with $F\propto(1/{t_f})^{1.5}e^{b}$. The tunneling-out time $t_{0}$ influences the tunneling amplitude $e^{b}$ and the excursion time $t_{f}$ determines the quantum diffusion effect $(1/{t_f})^{1.5}$ of the rescattering electron wave packet. These two factors work together to affect the amplitude of HHG electron trajectories. As mentioned above, the excursion time $t_{f}$ predicted by different models is comparable, so the difference between the amplitudes of HHG electron trajectories predicted by different models mainly depends on the difference between the predicted tunneling amplitudes $e^{b}$. Therefore, the amplitude of HHG electron trajectories obtained from TDSE provides a reliable numerical observation for qualitatively testing the tunneling-out time predicted by different models.
	
	It is also worth noting that in Fig. \ref{fig:g3}, the TRCM predictions of the tunneling-out time $t_0$ for both long and short trajectories are near to the laser peak time, so the amplitudes $e^{b}$ of the tunneling-electron wave packet generated at $t_0$ are also comparable for long and short trajectories. However, the excursion time $t_f$ for the long trajectory is remarkably larger than the short-trajectory one, implying that the quantum diffusion effect $(1/{t_f})^{1.5}$ for the long trajectory is stronger than the short-trajectory one. As a result, the TRCM predictions of the HHG amplitude $F\propto(1/{t_f})^{1.5}e^{b}$ for short trajectory are larger than the long-trajectory one, in agreement with the TDSE results seen in Fig. \ref{fig:g2}. By contrast, the SFA predicts that the long-trajectory HHG amplitude is remarkably larger than the short one.   
	
	\begin{figure}[t]
		\begin{center}
			\rotatebox{0}{\resizebox *{8.5cm}{8cm} {\includegraphics {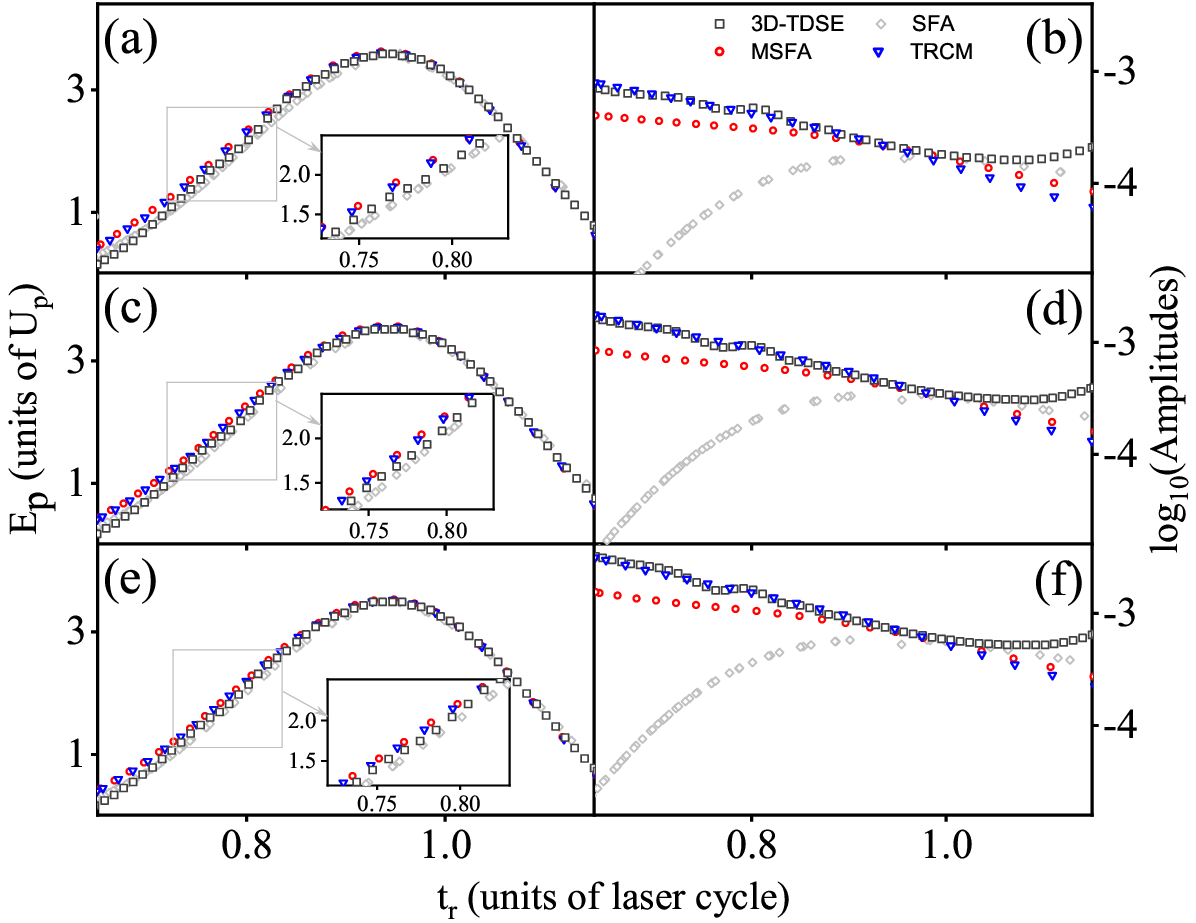}}}
		\end{center}
		\caption{Comparisons of HHG electron trajectories for the H atom, obtained with 3D-TDSE, TRCM, MSFA and SFA at $I=1.5\times10^{14}$W/cm$^{2}$ (a,b), $I=1.75\times10^{14}$W/cm$^{2}$ (c,d), and $I=2\times10^{14}$W/cm$^{2}$ (e,f).  Left column: return time $t_{r}$ versus return energy $E_p$. Right Column: return time $t_{r}$ versus amplitudes of HHG electron trajectories. For comparison, in each panel in the right column, the amplitude curves of TDSE, MSFA and SFA are vertically shifted to coincide with the amplitude curves of TRCM at $t_r=1T$. The laser wavelength used is $\lambda=900$ nm. For clarity, some enlarged results are shown in the insets.}
		\label{fig:g4}
	\end{figure}
	
	\begin{figure}[t]
		\begin{center}
			\rotatebox{0}{\resizebox *{8.5cm}{8cm} {\includegraphics {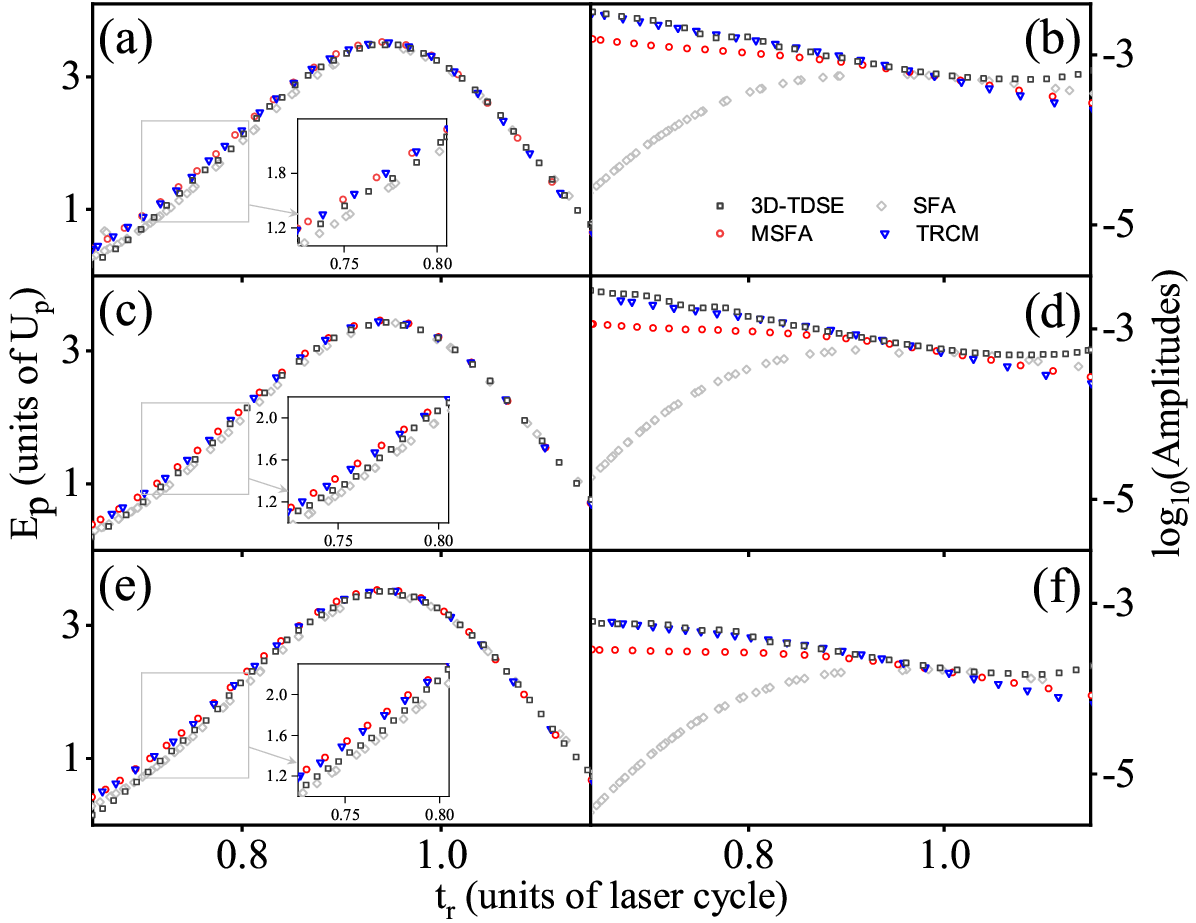}}}
		\end{center}
		\caption{Same as Fig. \ref{fig:g4} but for other laser parameters. The laser parameters used are $\lambda=800$ nm with $I=2.0\times10^{14}$W/cm$^{2}$ in (a,b), $\lambda=1000$ nm with $I=2.0\times10^{14}$W/cm$^{2}$ in (c,d) and $\lambda=1000$ nm with $I=1.5\times10^{14}$W/cm$^{2}$ in (e,f). }
		\label{fig:g5}
	\end{figure}

	In Fig. \ref{fig:g4}, we compare return time and amplitude of HHG electron trajectories obtained through TDSE and theoretical models at different laser intensities with $\lambda=900$ nm. The results in the left column of Fig. \ref{fig:g4} for varied laser intensities are similar. They indicate that the return times $t_{r}$ predicted by TRCM and MSFA are almost the same, somewhat earlier than those predicted by SFA, and closer to the TDSE results on the whole. The difference between TRCM and MSFA predictions and TDSE results is more significant only when the return energy $E_p$ of the rescattering electron along the short trajectory is lower. The above results can be understood as follows. Firstly, the long-range Coulomb potential provides a force to accelerate the rescattering electron to return to the nucleus in the recombination process, resulting an earlier return time $t_{r}$ than that predicted by SFA without considering the Coulomb. Secondly, for cases of short-trajectory electrons with lower return energy, the excursion distance is shorter and the Coulomb focusing \cite{Brabec} can play a more important role here. The Coulomb focusing effect was not fully considered in MSFA and TRCM. In order to better evaluate the Coulomb focusing effect, it may be necessary to consider more accurate continuous wave functions for continuum electrons with lower energy in the theoretical model.  
	
	Next, we turn to the theoretical predictions of amplitudes of HHG electron trajectories, as shown in the right column of Fig. \ref{fig:g4}. The amplitude curves of TDSE, MSFA and SFA shown here have been shifted by a vertical scaling factor so that the amplitudes of these curves coincide with the amplitude of TRCM at the return time of $t_r=1T$. For the long trajectory part with $t_r\geq0.9T$, the amplitude results of TRCM, MSFA and SFA are near to each other, and they are also comparable to the results of TDSE at higher energy. For lower energy, the TDSE amplitudes along the long trajectory differ from the model predictions. The potential reason may be that the emission of near-threshold harmonics related to the long trajectory is very complex, in which transitions between laser dressed states as well as multiple returns can also play an important role \cite{R H XU}. By contrast, for the short-trajectory part with $t_r<0.9T$, the model results differ remarkably from each other, especially for trajectories with lower energy. For example, at $t_r\approx0.65T$, the prediction of MSFA is almost two orders of magnitude higher than SFA and the prediction of TRCM is also several times higher than MSFA. In particular, in comparison with TDSE, the curve of TRCM for short trajectory basically agrees with the TDSE one, indicating similar scaling laws of TDSE and TRCM for the dependence of the HHG amplitude on the return time. By comparison, the scaling laws predicted by MSFA and SFA for short trajectory differ remarkably from TDSE. As the scaling law reflects the most fundamental dynamical property of the HHG process, the above analysis shows that there is a fundamental difference between TRCM and MSFA, and the TRCM developed here is capable of giving a quantitative description for HHG electron trajectory. Furthermore, the TRCM provides a clear physics picture for the significant increase of the short-trajectory amplitude in comparison with SFA. As discussed in Fig. \ref{fig:g3}, this significant increase can be attributed to the Coulomb induced initial velocity at the tunnel exit, which prevents tunneling electrons from immediately escaping, allowing electrons that tunnel out of the potential barrier to stay near the atomic nucleus for a small period of time.
	
	To further check our results, we have also performed comparisons for other laser parameters and relevant results are shown in Fig. \ref{fig:g5}. Results in first and the second rows of Figs. \ref{fig:g5} are obtained for different laser wavelengths of $\lambda=800$ nm (the first row) and $\lambda=1000$ nm (the second row) at a fixed laser intensity of $I=2.0\times10^{14}$W/cm$^{2}$, while results in the third row of Fig. \ref{fig:g5} are obtained with the laser parameters of $\lambda=1000$ nm and $I=1.5\times10^{14}$W/cm$^{2}$. The comparisons between the return-time and amplitude results of HHG electron trajectory obtained by different methods in Fig. \ref{fig:g5} give similar conclusions as in Fig. \ref{fig:g4}, indicating that the developed TRCM is suitable for describing HHG over a wide range of laser parameters. 
	
	\section{Extended comparisons and applicability}
	
	\begin{figure}[t]
		\begin{center}
			\rotatebox{0}{\resizebox *{8.5cm}{8cm} {\includegraphics {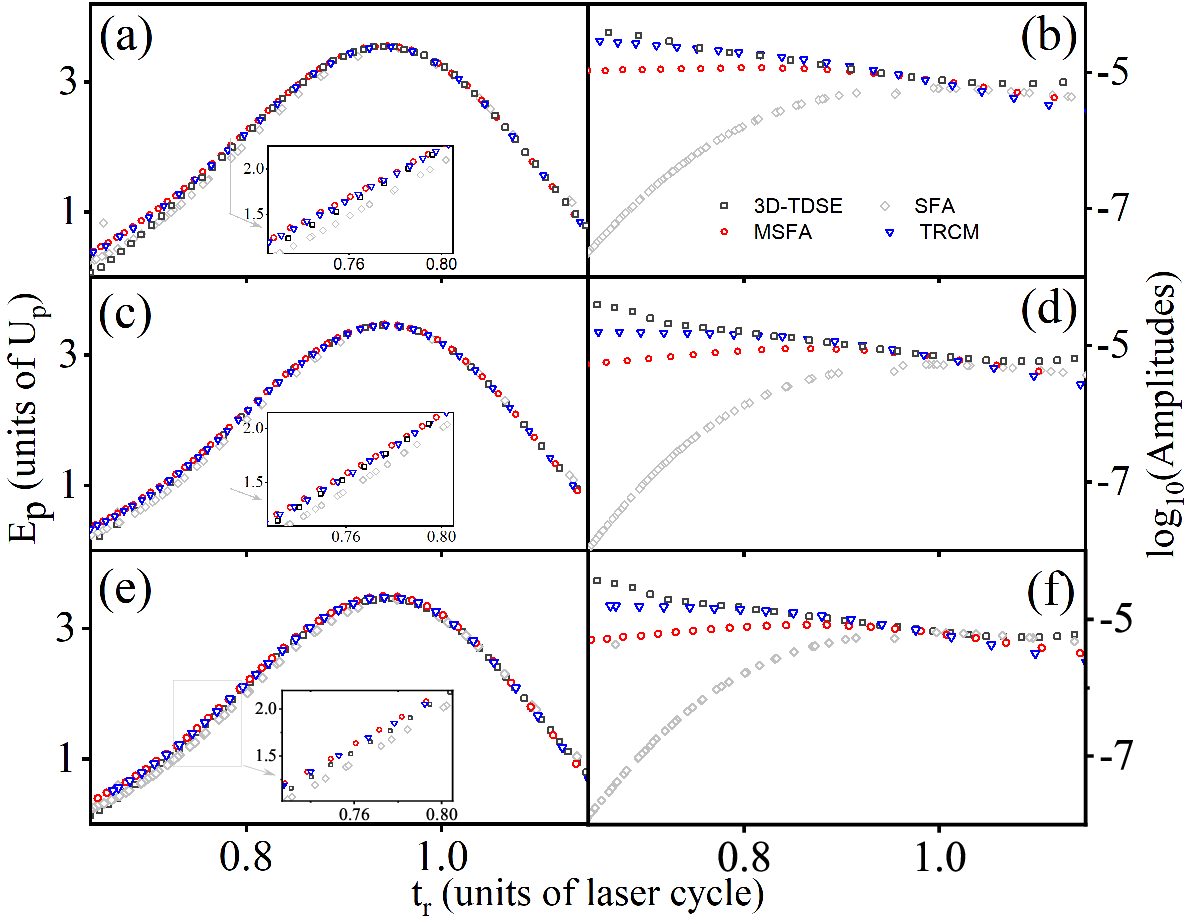}}}
		\end{center}
		\caption{Same as  Fig. \ref{fig:g4} but for the He atom at other laser parameters. The laser parameters used are $\lambda=700$ nm with $I=5.0\times10^{14}$W/cm$^{2}$ in (a,b), $\lambda=800$ nm with $I=5.0\times10^{14}$W/cm$^{2}$ in (c,d) and $\lambda=800$ nm with $I=4.0\times10^{14}$W/cm$^{2}$ in (e,f). }
		\label{fig:g6}
	\end{figure}

	We have also extended our simulations to the He atom with the 3D TDSE procedure introduced in Sec. II. A, at different laser intensities and wavelengths. In our calculations with single-active electron approximation, we use the model potential $V(\mathbf{r})=-(Z+a_1e^{-a_2r}+a_3re^{-a_4r}+a_5e^{-a_6r})/r$ proposed in \cite{Tong2005} to simulate the effective potential of He. Here, $Z=1$ is the effective charge and the parameters of $a_i$ are the fitting parameters with the values of $a_1=1.231$, $a_2=0.662$, $a_3=-1.325$, $a_4=1.236$, $a_5=-0.231$ and $a_6=0.48$ as shown in Table 1 in \cite{Tong2005}. Some typical results are shown in Fig. \ref{fig:g6}. Here, we compare the return-time and amplitude results of HHG electron trajectory obtained with different methods of TDSE, TRCM, MSFA and SFA, as in Fig. \ref{fig:g4} and Fig. \ref{fig:g5}. It can be observed from Fig. \ref{fig:g6}, that the results of return time and HHG amplitude obtained by different methods for He are similar to the cases for H. Specifically, the return-time results of MSFA and TRCM are similar and consistent with the TDSE results for relatively higher energy. Especially, when the return energy is not very low, the amplitude results of TRCM and TDSE are still very near to each other, showing similar scaling laws for the dependence of harmonic amplitude on the return time. By comparison, the predictions of MSFA for short-trajectory amplitudes are remarkably different from the TDSE ones, showing a different scaling law than TDSE, but significantly better than SFA . These results further support the applicability of the developed TRCM model for quantitatively describing HHG. We mention that compared with the HHG results of H in in Fig. \ref{fig:g4} and Fig. \ref{fig:g5}, the consistency between the TDSE and TRCM results of He in Fig. \ref{fig:g6} is slightly worse than that of H. The potential reason may be that the Coulomb potential used in TDSE simulations for He is more complex than that of H.	
	
	\begin{figure}[t]
		\begin{center}
			\rotatebox{0}{\resizebox *{8.5cm}{8cm} {\includegraphics {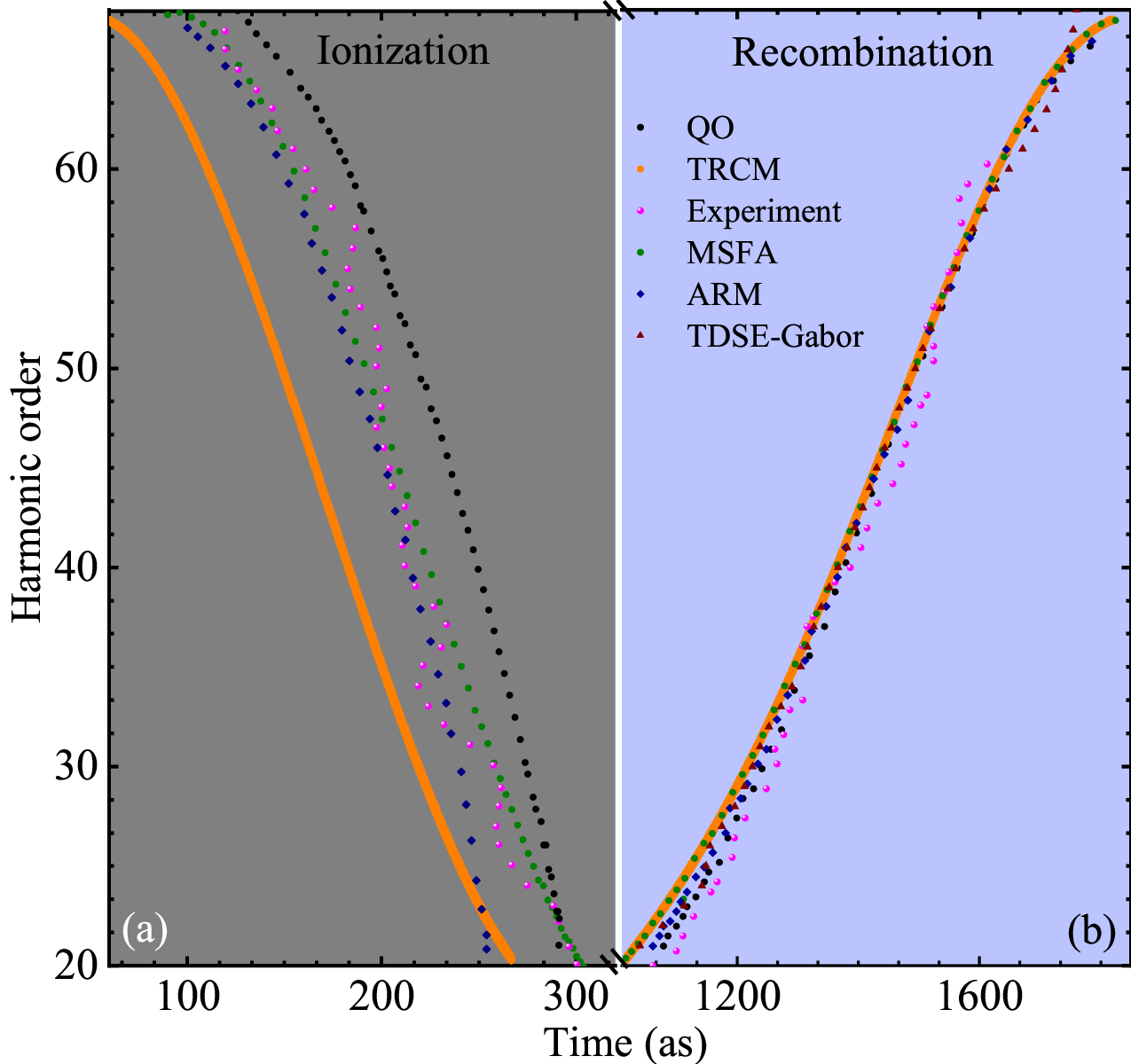}}}
		\end{center}
		\caption{Tunneling-out (ionization) (a) and return (recombination) (b) times of HHG electron trajectories obtained with SFA (i.e., QO, black circle), ARM (blue diamond), reconstruction from experimental data (pink spherical), TRCM (yellow line), MSFA (green circle), and Gabor analysis of 2D TDSE results (brown triangle) for the He atom at $\lambda=800$ nm and $I\approx3.8\times10^{14}$W/cm$^{2}$. The reconstruction results are taken from Ref. \cite{M. Y. Ivanov} and the ARM results are taken from Ref. \cite{L. Torlina}. The 2D TDSE results presented here are obtained through Gabor analysis \cite{M. Lein2} of the TDSE dipole acceleration calculated using the spectrum method \cite{M.D Feit}. The parameters of the Coulomb potential used in TDSE, TRCM and MSFA calculations are the same as in Ref. \cite{Hongchuan Du}. In TRCM, the tunneling-out time $t_0$ is not the ionization time $t_i$ and there is a lag $\tau$ between $t_i$ and $t_0$. Here, ${n_{f}}=2$ is used in Eq. (12) of TRCM to match 2D TDSE simulations.}
		\label{fig:g7}
	\end{figure}
	
	To show the differences between TRCM predictions and other results, we further compared TRCM predictions of short HHG electron trajectories triggered within half laser cycles with other models and experimental reconstructions. Relevant results are shown in Fig. \ref{fig:g7}, where we plot the tunneling-out time (a) and the return time (b) of short HHG trajectories for He atom, obtained by different theory models (including SFA, MSFA, TRCM and ARM \cite{L. Torlina}), two-dimensional (2D) TDSE and experimental reconstruction \cite{M. Y. Ivanov}. The results indicate that for the case of return time in Fig. \ref{fig:g7}(b), the predictions of MSFA, TRCM and ARM are very close to each other and are also consistent with the TDSE results. They are 10 to 20 attoseconds earlier than the SFA predictions and the experimental reconstruction, especially when the return energy is low. In contrast, the tunneling-out times obtained by different methods differ remarkably from each other, as seen in Fig. \ref{fig:g7}(a). Overall, the TRCM predictions for the tunneling-out time are 40 (80) attoseconds earlier than the results of MSFA (SFA). The predictions of ARM for higher and lower energy are near to the results of MSFA and TRCM, respectively, while the results of experimental reconstruction seem to be distributed between the results of ARM and SFA and fluctuate around the results of MSFA. It should be mentioned that the experimental reconstruction results are obtained by ingeniously constructing SFA-based equations to deduce the time information from experimental measurements. From the amplitude results in Fig. 4 to Fig. 6, experiments with measuring the yields of harmonics related only to short or long HHG electron trajectories \cite{Lewenstein1996} are expected to provide experimental data to directly test the accuracy of theoretical models for HHG electron trajectories.

	Before conclusion, we give further discussions on the applicability of the TRCM model developed here, which is used to describe the effects of Coulomb potential on electron trajectories of HHG. Firstly, this model developed here uses the general TRCM theory for ATI \cite{xie3,Huang2023} to describe the Coulomb effect in the tunneling ionization process of HHG. Therefore, the applicability of this model is partly dependent of the applicability of the general TRCM. The latter is more applicable for longer laser wavelengths, higher laser ellipticities and intermediate laser intensities with moderate ionization yields \cite{chen2025}. Secondly, this model developed here uses the Newton equation to describe the Coulomb effect in the recombination process of HHG. Therefore, the return time evaluated by the model is real. This is generally true for the Keldysh parameter \cite{Keldysh} $\gamma$ being smaller than the unity, i.e., $\gamma=\omega_0\sqrt{2I_p}/{E_{0}}<1$. For $\gamma\geq1$, an imaginary component of $t_r$ may be necessary for the rescattering electron to return to the nucleus, which can not be described by the present model. For the results in Fig. \ref{fig:g4} and Fig. \ref{fig:g5}, the Keldysh parameter decreases with increasing the laser intensity and wavelength. By calculating the difference between relevant curves obtained by different methods in Fig. \ref{fig:g4} and Fig. \ref{fig:g5}, we observe that the return-time difference between TDSE and TRCM along the short trajectory with $E_p\geq1U_p$ decreases from an average of about 30 attoseconds in Fig. \ref{fig:g4}(a) to about 15 attoseconds in \ref{fig:g5}(c) (also see the Appendix). In these two figures, compared to the return time difference, the amplitude difference between TDSE and TRCM along the short trajectory is small and is not sensitive to the Keldysh parameter, suggesting that the tunneling-out time predicted by TRCM is very near to the actual one in TDSE in a wide parameter region. Thirdly, this model developed here uses the tunneling amplitude $e^{b}$ (see Eq. (20)) to describe the HHG amplitude as in \cite{xie2,xie1} and does not consider the recombination dipole matrix element \cite{M. Lewenstein2} describing continuum-bound transition. This may not be crucial for cases of atoms. However, for molecules, two-center interference described by the recombination dipole \cite{Lein2002,chen2009,sun2022} influences significantly on HHG, and therefore needs to be considered in the present model. In addition, for polar molecules, the permanent dipole also plays an important role in both ionization and recombination processes of HHG \cite{che2024} and therefore needs to be included in the present model.      
	
	\section{Conclusion}
	In summary, we have studied the effect of Coulomb potential on HHG from atoms in a wide range of laser parameters. We have proposed a Coulomb-included model called TRCM for quantitatively describing the tunneling-out time, the return time and the amplitude of HHG electron trajectories. This model considers that the Coulomb potential plays different roles in the ionization and recombination processes of HHG, therefore different methods are used to handle different effects. In the ionization process, the Coulomb-related symmetry dominates. That is to say, the system which is initially located in a quantum state with a high Coulomb-related symmetry tends to maintain this symmetry as much as possible before ionization. This effect can be described semiclassically by introducing a Coulomb-induced initial velocity at the tunnel exit, which prevents the tunneling electron from immediately escaping. In the recombination process, the Coulomb acceleration dominates. That is to say, when the rescattering electron is driven back to the atomic nucleus, the Coulomb field will classically accelerate it. This effect can be described by numerical solution of Newton equation including both electric force and Coulomb force.
	
	In comparison with the general SFA that neglects the Coulomb effect, the TRCM including the Coulomb effect predicts a time lag $\tau$ between the time $t_{0}$ at which the electron tunnels out of the barrier and the time $t_{i}$ at which the electron is free. Due to this lag, for a harmonic with a specific energy $\Omega$, the tunneling-out time $t_{0}$ in TRCM is about 100 attoseconds earlier than that in SFA. This effect leads to a significant increase in HHG amplitude for short trajectories, making the amplitude of short trajectories larger than that of long trajectories, which is consistent with TDSE simulations. This effect also allows the electron trajectories triggered by the rising edge of the laser field to contribute to HHG, significantly improving the efficiency of HHG. In addition, in comparison with the MSFA that considers the Coulomb effect in both ionization and recombination processes through numerical solution of Newton equation, the TRCM considers the Coulomb effect in the ionization process through a semiclassical and analytical approach and is able to give a clear physical picture for how the Coulomb potential influences the tunneling-ionization process of HHG. In particular, the TRCM is also able to provide the scaling law for the dependence of HHG amplitude on return time that is consistent with the TDSE simulation, while the scaling law predicted by the MSFA differs remarkably from TDSE. Due to the fact that the scaling law reflects the most fundamental dynamical characteristics of the HHG process, this consistency between TDSE and TRCM indicates the applicability of the TRCM developed here in quantitatively describing the HHG electron trajectories. These results provide deeper insights into the HHG mechanism and give suggestions on designing procedures to enhance the attosecond pulse intensity obtained through HHG. Since the TRCM model developed here considers the Coulomb effect in both the tunneling and the rescattering processes, in addition to HHG, it can also be used to study strong-field ionization significantly affected by rescattering.
	
	This work was supported by the National Natural Science Foundation of China (Grant Nos. 12404330, 12304303, 12347165, 12174239, and 12574376), Shaanxi Fundamental Science Research Project for Mathematics and Physics (Grant No. 23JSY022), Hebei Natural Science Foundation (Grant No. A2022205002), and Science and Technology Project of Hebei Education Department (Grant No.QN2022143).
	
	\appendix*
	
	\section{Validity of TRCM for ionization and significance of TRCM for HHG}
	
	The TRCM model is first developed to quantitatively describe strong-field ionization. It is based on SFA but includes the Coulomb effect. The TRCM considers that the effect of the Coulomb potential near the atomic nucleus plays an important role in tunneling, and after tunneling, the Coulomb effect can be neglected in subsequent processes. However, for the rescattering process which is a basic step in HHG, the ionized electron will return to the atomic nucleus, and the Coulomb potential will certainly have an important impact. Therefore, the TRCM first developed is more applicable for cases where the rescattering process plays a small role such as in the case of attoclock. In practice, the TRCM is able to quantitatively reproduce a series of recent attoclcok experimental curves, providing a consistent explanation for these experiments. On the other hand, a quantitative model for describing HHG electron trajectories is crucial for HHG-based attosecond measurements. Since ionization is the first step of HHG and the TRCM is capable of providing a quantitative description for ionization, in the paper, through appropriately considering the Coulomb effect in the rescattering process, we further develop the TRCM to quantitatively describe the HHG. Next, we give a detailed discussion on the validity of the general TRCM model for ionization and the significance of the TRCM model developed here for HHG.

	\subsection{Coulomb-included orbit-based strong-field models for ionization}
	
	\textit{Limitations of SFA.} Strong-field approximations (SFA) along with the electron-trajectory theory \cite{M. Lewenstein,M. Lewenstein2,M. Lewenstein3} relating to the saddle-point solution of SFA for ionization and HHG have been widely used in understanding the strong-field phenomena and probing the electron dynamics. The SFA ignores the contributions of Coulomb potential and higher bound eigenstates, so it is basically only accurate for delta-potential atoms with only one bound eigenstate. For quantitative descriptions of strong-field processes which are essential in attosecond-resolved measurements, the inclusion of the Coulomb effect into the SFA is necessary. 
	
	\textit{Main ideas of general Coulomb-included orbit-based models.} Because the analytical solution of the Schrödinger equation for a strong-laser-driven atom system with the long-range Coulomb potential is very difficult to obtain at present, Coulomb-included orbit-based strong-field models, such as introduced in \cite{xie2,xie1,Goreslavski2004,Yan2010,Li2014,Lai2015,Shvetsov2016,Bray2018,Peng2025} have been developed to overcome this difficulty. The main calculation steps of these models are as follows. The SFA saddle-point equation is first solved to obtain the electron trajectories including the initial velocity (i.e., the exit velocity) $\textbf{v}(t_0)$, the initial position (i.e., the exit position or the tunnel exit) $\textbf{r}(t_0)$ at the tunneling-out time $t_0$. Then these initial conditions are used in the solution of the Newton equation including the electric force and the Coulomb force. The evolution of the Newton equation gives the Coulomb-modified drift momentum $\textbf{p}'$ of the photoelectron. Based on the calculation steps, the main ideas of these models can be summarized as follows. 1) The Coulomb potential after the tunnel exit (which we call the far-nucleus Coulomb potential) plays an important role in the evolution of the system and can be treated classically, while the Coulomb potential before the tunnel exit (the near-nucleus Coulomb potential) plays a small role and can be neglected. 2) At the tunnel exit, due to the inclusion of the Coulomb potential, the whole energy of the tunneling electron is generally negative. From a semiclassical perspective, this implies that the tunneling electron is located at a quasibound state. 3) The classical Coulomb effect can be attributed to a Coulomb-induced initial velocity which is contrary to the direction of the exit position and prevents the tunneling electron from escaping \cite{Goreslavski2004}, while Coulomb-related velocities in other directions are neglected. These models have successfully explained many important strong-field phenomena and have been widely used in strong-field physics. However, these models have some limitations, mainly in terms of quantification. One intuitive reason for these limitations is the neglect of the near-nucleus Coulomb potential. Intuitively, the near-nucleus Coulomb potential should play a more important role in the evolution of the system than the far-nucleus one.  
	
	\subsection{TRCM model for ionization}
	\textit{Characteristics of near-nucleus Coulomb potential.} When the far-nucleus Coulomb effect can be treated using CLASSICAL methods, the near-nucleus Coulomb effect has to be treated using QUANTUM mechanical methods, as quantum effects dominate near the atomic nucleus. A direct method to address this quantum issue is to include higher bound eigenstates in SFA in addition to the ground state. However, in this case, the SFA is also difficult to solve, as it is well known that even for a two-level atom including two bound eigenstates in a strong low-frequency laser field, the analytical solution is difficult to obtain. On the other hand, it is well known that the bound eigenstate of a hydrogen-like atom agrees with the virial theorem. This theorem also approximately holds for a bound electronic wave packet formed in laser-induced tunneling ionization of the hydrogen-like atom, which is composed of coherent superposition of a series of higher bound eigenstates of the field-free atom system, indicating that the Coulomb-related symmetry plays an important role in the evolution of the laser-driven atom system  \cite{chen2025,Wu2025}. 
	
	\textit{Main ideas of TRCM.} Inspired by the Coulomb-related symmetry in tunneling ionization of atoms, the TRCM bypasses the difficulty of solving the time-dependent Schrödinger equation of a multiple-level system from the perspective of symmetry of an atom system \cite{xie3,Huang2023,Jia-Nan Wu,chen2025}. Specifically, The TRCM model follows the basic ideas of other Coulomb-included orbit-based strong-field models but considers the near-nucleus Coulomb effect. It assumes that at the tunnel exit, the tunneling electron is still located at a quasibound state, as performed in other models. This quasibound state is composed of higher bound eigenstates of the atom and approximately agrees with the virial theorem. With the help of a strong laser field, a small portion of the quasi-bound state can be ionized, and the ionized portion can be regarded as a classical particle with a Coulomb-induced velocity at the tunnel exit (which is also called the Coulomb-corrected initial velocity here). The amplitude of this velocity is given by the virial theorem, and similar to other models, the direction of this velocity is opposite to the tunneling direction. In other words, the ONLY IMPORTANT DIFFERENCE between TRCM and other Coulomb-included orbit-based models is the different description of the Coulomb-induced velocity amplitude at the tunnel exit. 
	
	\textit{Different evaluations of Coulomb-induced velocity amplitude.} The expression of the Coulomb-modified momentum $\textbf{p}'$ related to the Coulomb-free SFA momentum $\textbf{p}$ at the tunneling-out time $t_0$ (i.e., the real part of the saddle-point time $t_s$) in TRCM can be written as $\textbf{p}'=\textbf{v}(t_0)+\textbf{v}_i-\textbf{A}(t_0)$ (Eq. (14) in the paper). In the above expression, the term $\textbf{v}(t_0)=\textbf{p}+\textbf{A}(t_0)$ is the tunneling-related Coulomb-free initial velocity. It arises from the classic-quantum correspondence and is generally considered to reflect the nonadiabatic effect  in tunneling ionization \cite{Yan2010,Boge,Huang2023}. When the Keldysh parameter is small, this term can be neglected in Eq. (14), then one can also obtain the adiabatic version of TRCM, which is useful for analytically deducing the scaling law \cite{Huang2023,chen2025}. The term $\textbf{v}_i$ is the Coulomb-induced velocity at the tunnel exit and the term $-\textbf{A}(t_0)$ is the velocity induced by the laser field. In other models, the amplitude of the Coulomb-induced velocity $\textbf{v}_i$ can be approximately evaluated with the perturbation theory (PT) \cite{Goreslavski2004} or the Keldysh-Rutherford (KR) model  \cite{Bray2018,Peng2025}, which puts an emphasis on the classical properties of the FAR-NUCLEUS Coulomb effect. In TRCM, the amplitude of $\textbf{v}_i$ is approximately evaluated with the virial theorem which gives the average velocity (root mean square velocity related to the average kinetic energy) of the bound electronic wave packet contributing to tunneling and puts an emphasis on the quantum properties of the NEAR-NUCLEUS Coulomb effect. The expression for the velocity amplitude $v_i$ in TRCM is ${v}_{i}=\sqrt{\left|V(\textbf{r}_0)\right|/{n_{f}}}$ (Eq. (12) in the paper). Here, $V(\textbf{r})$ is the Coulomb potential, $\textbf{r}_0$ is the exit position and the parameter $n_f$ is the dimension of the system studied. For actual three-dimensional cases, we have $n_f=3$, and for two-dimensional TDSE simulations frequently performed in theory studies, we have $n_f=2$. The above expression with the parameter $n_f$ implies that only a portion of the average kinetic energy of the bound electronic wave packet associated with tunneling contributes to the velocity amplitude $\textbf{v}_i$. The applicability of these expressions of Eq. (12) and Eq. (14) in TRCM has been explored in \cite{chen2025}, where related issues such as “Transverse momentum” and “Classical and quantum virial theorem”, etc., have also been discussed. This applicability can be further checked by COMPARING predictions of TRCM with experiments and predictions of other theories. Relevant comparisons are shown in Fig. S1 and Fig. S2. 
	
	\subsection{Comparisons between predictions of TRCM and other models and experiments}
	In \cite{xie3,cheaps,Huang2023}, we have compared the predictions of TRCM with a series of attoclock experiments \cite{Sainadh,Boge,Landsman,Eckle3,Quan,Torlina2015,Landsman2013}. Here, we further compare the predictions of TRCM with those of other orbit-based models such as PT, KR and MSFA. 
	
	\textit{Cases of different laser intensities.} In Fig. S1, we show the attoclock offset angle $\theta$ measured in experiments and predicted by different theory models of TRCM, PT, KR and MSFA for different targets at different laser parameters. From the results in Fig. S1, it can be observed that for different targets, the TRCM predictions quantitatively agree with the experimental curves with the change of laser intensity. By comparison, the predictions of other models differ remarkably from the experimental results on the whole. In addition, in comparison with the results of MSFA and PT, the predictions of KR are nearer to the experimental results on the whole. The results of the PT which can be considered as a direct approximation of the MSFA are similar to the MSFA results, but with a vertical shift. This similarity implies that the PT holds the essence of MSFA, and this shift means that the dynamics of the electron after tunneling described by MSFA is somewhat more complex than PT. The more important difference between these models of TRCM, PT, KR and MSFA is the different scaling law of the predicted offset angle relative to the laser intensity. The experimental curves of different targets in Fig. S1 all show a decreasing trend with the increase of laser intensity. By comparison, for cases of higher laser intensities, the curves of MSFA and PT are  insensitive to the laser intensity. The curves of TRCM and KR also show a decreasing trend with the increase of laser intensity, and the curve of KR decreases faster than TRCM. In fact, according to the relation of $\tan\theta\approx\theta\approx{p}'_x/{p}'_y$, when the Keldysh parameter is smaller than the unity, namely $\gamma=\omega\sqrt{2I_p(1+\epsilon^{2})}/{E_{L}}<1$, the term $\textbf{v}(t_0)$ in the expression of $\textbf{p}'$ can be neglected and the scaling law of the offset angle with respect to the laser parameters can be written as    
	\begin{equation}
		\theta\sim \omega(1+{\epsilon}^{2})^{1/4}{\epsilon}^{-1}{E_L}^{-1/2}{I_p}^{-1/4}
	\end{equation}
	for TRCM, and it is 
	\begin{equation}
		\theta\sim\omega^2(1+{\epsilon}^2)^{1/2}{\epsilon}^{-2}{E_{L}}^{-1}{I_p}^{-1/2}	\end{equation}
	for KR and 
	\begin{equation}
		\theta\sim \omega{\epsilon}^{-1}{I_p}^{-1}	
	\end{equation}
	for PT. Here, $E_L$ is the laser amplitude corresponding to the peak intensity of the laser pulse, $\omega$ is the laser frequency, $\epsilon$ is the laser ellipticity and $I_p$ is the ionization potential of the target. In obtaining the above scaling relations, similar derivations as in \cite{Huang2023} have been used. The relations clearly show that with the increase of the laser intensity $E_L$, the offset angle in KR decreases faster than that in TRCM and the offset angle in PT is independent of the laser intensity. When the TRCM prediction for the scaling law of $\theta$ relative to the laser intensity is more consistent with the experimental result than the KR prediction, the predictions of PT and MSFA are remarkably inconsistent with the experimental result.  
	
	\setcounter{figure}{0}
	\renewcommand{\thefigure}{S\arabic{figure}}
	\captionsetup[figure]{name=Figure}
	\begin{figure}[t]
		\begin{center}
			{\resizebox *{8.5cm}{8cm} {\includegraphics {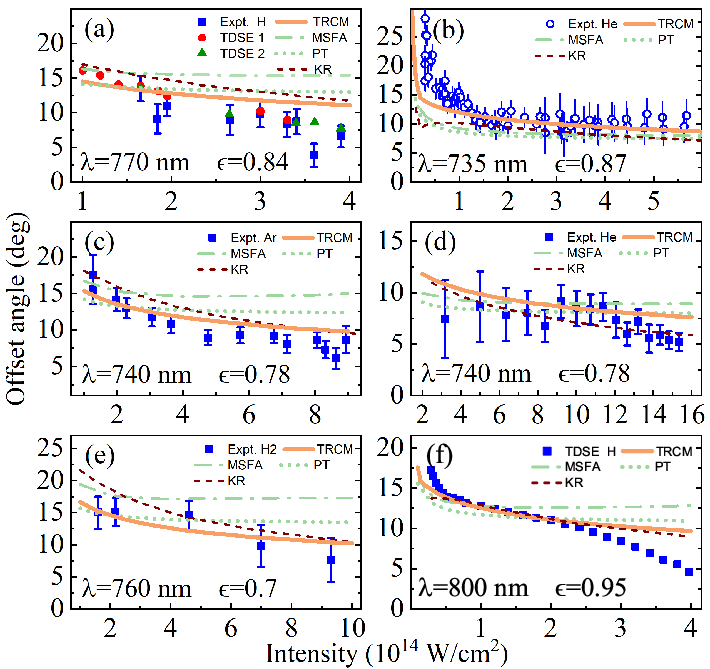}}}
		\end{center}
		\caption{Comparisons of predictions of TRCM and other models of KR, PT and  MSFA for the attoclock offset angle $\theta$ with experimental results obtained for different targets at different laser parameters. The offset angle $\theta$ is defined by the Coulomb-modified drift momentum $\textbf{p}'$, which is related to the most probable route (MPR) of the electron trajectory with the maximal amplitude, expressed as $\tan\theta={p}'_x/{p}'_y$ \cite{xie3,Huang2023,cheaps}. The results of all TRCM, PT and KR are obtained with the expression of $\textbf{p}'=\textbf{v}(t_0)+\textbf{v}_i-\textbf{A}(t_0)$ of Eq. (14) in the paper for this momentum $\textbf{p}'$ related to the MPR.  The difference between TRCM, PT and KR is that the amplitude of $\textbf{v}_i$ in Eq. (14) is obtained with the virial theorem (i.e., the expression of Eq. (12) in the paper, namely  ${v}_{i}=\sqrt{Z/(3r_0)}$) for TRCM, the expression of Eq. (8) in \cite{Goreslavski2004} multiplied by the effective charge $Z=\sqrt{2I_p}$ for PT (namely $v_i=Z\pi \sqrt{2r_0/\left|\textbf{E}(t_0)\right|}/(4{r_0}^2)$) and the expression of Eq. (5) in \cite{Peng2025} for KR (namely $v_i=2Z/(pr_0)$). The MSFA results are obtained with the numerical solution of Eq. (7) in the paper which gives the Coulomb-modified drift momentum $\textbf{p}'=\dot{\textbf{r}}(\textbf{p},t\rightarrow\infty)$. The PT result can be considered as a direct analytical approximation of the MSFA result. The experimental and TDSE data are taken from: (a) Ref. \cite{Sainadh}; (b) Refs. \cite{Boge,Landsman}; (c) and (d) Ref. \cite{Eckle3}; (e) Ref. \cite{Quan}; (f) Ref. \cite{Torlina2015}. The laser parameters used in calculations are as shown. }
		\label{fig:gs1}
	\end{figure}	
	
	\setcounter{figure}{1}
	\renewcommand{\thefigure}{S\arabic{figure}}
	\captionsetup[figure]{name=Figure}
	\begin{figure}[t]
		\begin{center}
			\rotatebox{0}{\resizebox *{8.5cm}{7cm} {\includegraphics {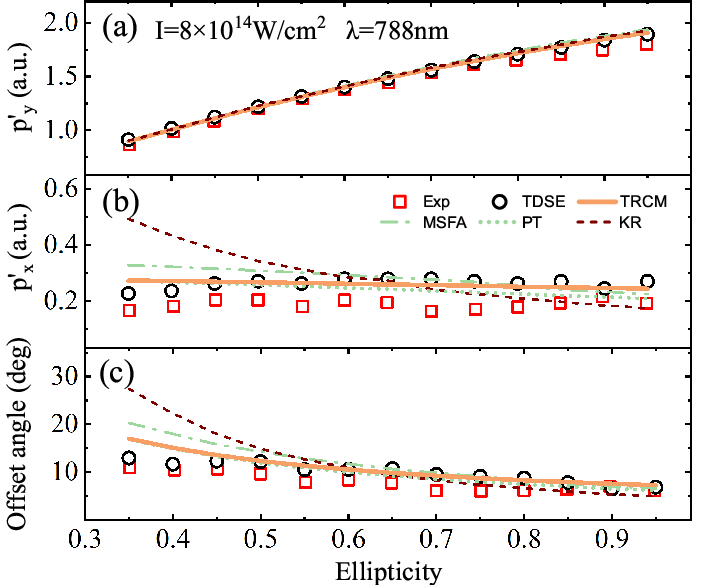}}}
		\end{center}
		\caption{Comparisons of predictions of TRCM and other models of KR, PT and  MSFA for the attoclock offset angle $\theta$ and the Coulomb-modified drift momentum (${p}'_x$, ${p}'_y$) related to the MPR with experimental results obtained for the He atom at different laser ellipticities. The calculation details of TRCM, PT, KR and MSFA for the offset angle and the drift momentum related to the MPR are as introduced in Fig. S1. In (a)-(c), the experimental data are taken from \cite{Landsman2013} and the TDSE data are taken from \cite{Huang2023}. The laser parameters used in calculations are as shown.}
		\label{fig:gs2}
	\end{figure}

	\textit{Cases of different laser ellipticities.} The applicability of TRCM can also be observed in Fig. S2 where we compare the predictions of different models for the attoclock observables of the offset angle $\theta$, and the Coulomb-modified momentum (${p}'_x$, ${p}'_y$) related to the MPR with experimental results obtained at different laser ellipticities for the He atom. With similar derivations as in \cite{Huang2023}, the scaling law of the momentum component ${p}'_x$ relative to the laser and atomic parameters is 
	\begin{equation}
		{p}'_x\sim (1+{\epsilon}^{2})^{-1/4}{E_L}^{1/2}{I_p}^{-1/4}
	\end{equation}
	for TRCM, and it is 
	\begin{equation}
		{p}'_x\sim\omega{\epsilon}^{-1}{I_p}^{-1/2}
	\end{equation}
	for KR and 
	\begin{equation}
		{p}'_x\sim (1+{\epsilon}^{2})^{-1/2}{E_L}{I_p}^{-1}
	\end{equation}
	for PT. The predictions of these models for ${p}'_y$ are similar. From the results in Fig. S2, one can observe that the scaling laws of $\theta$ and ${p}'_x$ relative to the laser ellipticity predicted by the TRCM still agree well with the experimental curves. By comparison, the predictions of the KR differ remarkably from the experimental results for cases of smaller ellipticities. The predictions of the PT, which is an analytical approximation of the MSFA, are comparable with the TRCM predictions here. However, the predictions of the MSFA itself also differ from the TRCM results, especially for cases of smaller ellipticities. For example, in Fig. S2(b), the experimental results for the momentum component  ${p}'_x$ are almost unchanged with the increase of ellipticity. In contrast, the KR predicts a remarkably decreasing trend for the momentum component when increasing the ellipticity. Therefore, for changing the laser intensity, the scaling laws of the experimental observables in attoclock predicted by PT and MSFA are remarkably inconsistent with the experimental curves, while for changing the laser ellipticity, the KR predictions of the scaling laws differ significantly from the experimental curves. In both cases, the TRCM provides the scaling laws in good agreement with the experimental measurements. The different scaling laws predicted by these models stem from the different properties of the Coulomb effect described in these models. In TRCM, the Coulomb effect is described using the virial theorem, which emphasizes the symmetry of the system from a quantum perspective. In PT and MSFA, the Coulomb effect is described by the Newton equation, which emphasizes the classical properties of the Coulomb field. In KR, the Coulomb effect is described by the Rutherford scattering formula, which also emphasizes the symmetry of the system to some extent, but from a classical perspective. Therefore, the Coulomb-induced velocities given by these models also differ essentially from each other, resulting in different scaling laws for attoclock observables. As the scaling laws reflect the MOST FUNDAMENTAL physical characteristics of the laser driven system, these results in Fig. S1 and Fig. S2 show the applicability and necessity of TRCM in describing strong-field ionization, as well as certain limitations of KR, PT and MSFA in describing strong-field ionization. 
	
	\textit{Potential mechanisms and properties of tunneling.} Based on the results in Fig. S1 and Fig. S2, which show the applicability of the TRCM model characterized by the expressions of Eq. (12) and Eq. (14), we can further discuss the potential mechanisms and properties of tunneling revealed by the TRCM. Firstly, the Coulomb-related space symmetry plays an important role in shaping the wave packet of the tunneling electron, which is further supported by numerical experiments \cite{chen2025}. So the virial theorem can be used to evaluate the average velocity (root mean square velocity related to the average kinetic energy) of the bound electronic wave packet contributing to tunneling in Eq. (12). Secondly, the bound electronic wave packet contributing to tunneling is composed of coherent superposition of higher bound eigenstates of the field-free system, and therefore has a small amplitude. The average velocity given by virial theorem corresponds to the main velocity component of the bound wave packet with a small amplitude, and therefore contributes mainly to tunneling. So the average velocity given by virial theorem can be used as the amplitude of the Coulomb-corrected velocity $\textbf{v}_i$ in Eq. (14). Thirdly, according to the saddle-point theory, tunneling ionization is a semiclassical process. From a classical perspective, in tunneling ionization, the electron escapes along the polarization direction of the laser field (tunneling direction). During the escape process, the Coulomb force will cause a velocity opposite to the tunneling direction \cite{Goreslavski2004}. Corresponding to the classical picture, from a quantum perspective, only the velocity component of the bound wave packet that not only has a velocity amplitude close to the average velocity, but also has a velocity direction opposite to the tunneling direction, will have a significant impact on tunneling. So the direction opposite to the tunneling direction can be taken as the direction of the Coulomb-corrected velocity $\textbf{v}_i$ in Eq. (14).
	
	\setcounter{figure}{2}
	\renewcommand{\thefigure}{S\arabic{figure}}
	\captionsetup[figure]{name=Figure}
	\begin{figure}[t]
		\begin{center}
			\rotatebox{0}{\resizebox *{8.5cm}{7cm} {\includegraphics {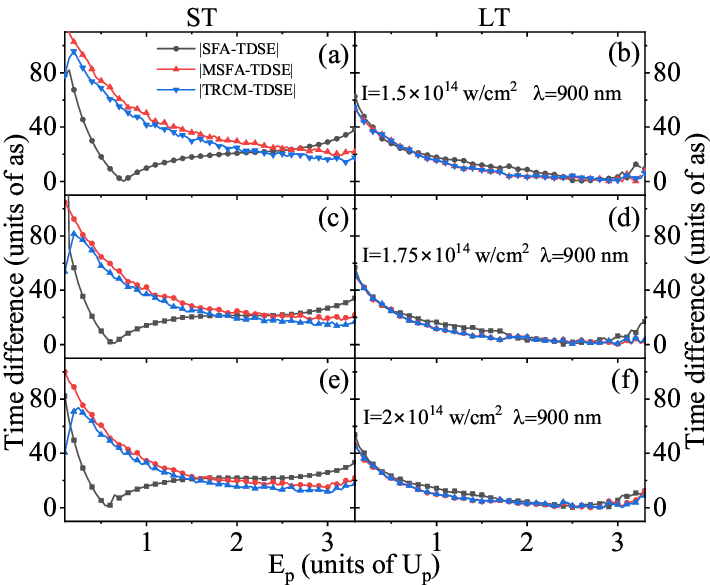}}}
		\end{center}
		\caption{Comparisons for the absolute value of difference between the return-time results of TDSE and theory models of SFA, MSFA and TRCM in the left column of Fig. 4 for short trajectory (ST, the left column) and long trajectory (LT, right). The laser parameters used are as shown.}
		\label{fig:gs3}
	\end{figure}

	\setcounter{figure}{3}
	\renewcommand{\thefigure}{S\arabic{figure}}
	\captionsetup[figure]{name=Figure}
	\begin{figure}[t]
		\begin{center}
			\rotatebox{0}{\resizebox *{8.5cm}{7cm} {\includegraphics {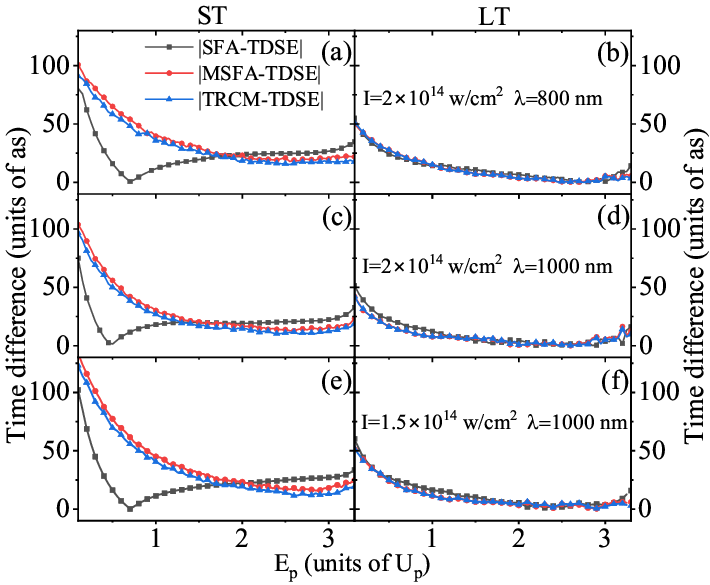}}}
		\end{center}
		\caption{Same as Fig. S3 but for the return-time results in the left column of Fig. 5.}
		\label{fig:gs4}
	\end{figure}

	\subsection{TRCM model for HHG}
	Since ionization is the first step of HHG, the basic limitation of MSFA in describing ionization discussed above also indicates the limitation of the MSFA-based method in describing HHG such as the method introduced in Ref. \cite{xie1} of the paper. On the other hand, a quantitative model describing the electron trajectories of HHG is crucial for constructing the probing procedure in HHG-based attosecond measurements, as performed in \cite{M. Y. Ivanov,Lein2013}. Therefore, in the present work, we develop the TRCM to quantitatively describe the HHG. To do so, one needs to further include the Coulomb effect in the rescattering process into the TRCM. Now, the question has become how to reasonably consider this effect. Intuitively, one can consider the use of Eq. (7) of MSFA in the paper but with the initial conditions given by Eqs. (18) and (19) of TRCM. However, this treatment is inapplicable, as the TRCM has considered the main Coulomb effect on ionization. The direct solution of Eq, (7) with Eqs. (18) and (19) of TRCM will OVERESTIMATE the Coulomb effect on ionization. To overcome this difficulty, Eq. (17) of the paper is proposed, which only includes the Coulomb effect after the rescattering electron begins to return to the nucleus. Equation (17) along with the initial conditions of Eqs. (18) and (19) are the main results of the paper, and the applicability of these expressions is explored in detail in the paper. 
	
	From a quantitative perspective, firstly, the tunneling-out times predicted by these expressions of TRCM are several tens of attosecond earlier than those predicted by the MSFA, as shown in Fig. 3(a). The prediction of TRCM for the tunneling-out time is supported by the observation that the scaling law for the dependence of harmonic amplitude on the return time predicted by these expressions of TRCM is consistent with TDSE, while the prediction of MSFA differs remarkably from TDSE, as seen in the right column of Fig. 4 to Fig. 6. These results are of significant importance in HHG-based attosecond measurements, as they may be crucial in quantitatively interpreting experimental results and extracting dynamic information of the system with high temporal resolution. Secondly, the return times predicted by these expressions of TRCM for short trajectories are also somewhat nearer to the TDSE results than the MSFA. To illustrate this point more clearly, in Fig. S3 and Fig. S4, we plot the differences between relevant curves in Fig. 4 and Fig. 5 to highlight the difference between predictions of different models. It can be observed that for long trajectories in the right column of Fig. S3 and Fig. S4, the predictions of TRCM and MSFA are very near to each other and are also somewhat nearer to the TDSE results than SFA on the whole. For short trajectories in the left column of Fig. S3 and Fig. S4, the difference between curves of TDSE and TRCM is somewhat smaller than that between TDSE and MSFA. It is worth noting that for short trajectories with lower energy (lower than 1$U_p$), the predictions of all models differ remarkably from the TDSE results, and the potential reason may be related to Coulomb focusing, as discussed in Fig. 4. While for short trajectories with higher energy (higher than 1$U_p$), the difference between curves of TRCM and TDSE becomes to decrease when the Keldysh parameter decreases, changing from an average of about 30 attoseconds in Fig. S3(a) to about 15 attoseconds in Fig. S4(c), as discussed in Sec. IV. 
	
	From a physical perspective, these expressions of TRCM clearly indicate the QUANTUM and CLASSIC effects of Coulomb potential in the ionization and recombination processes of HHG, which have not been reported in previous studies. In particular, in comparison with the MSFA which numerically treats the Coulomb effect in both ionization and recombination (i.e., rescattering) processes of HHG, the developed TRCM with these expressions is a semi-analytical model which treats the Coulomb effect analytically in the ionization process and numerically in the recombination process. So the developed TRCM is able to provide a clear physical picture for earlier tunneling-out times than predicted by the SFA. Therefore, the developed TRCM model characterized by these expressions may be a promising new tool for quantitatively studying HHG.

\end{document}